\documentclass[9pt]{elife}

\usepackage{lipsum} 
\usepackage[version=4]{mhchem}
\usepackage{siunitx}
\DeclareSIUnit\Molar{M}
\usepackage{amsmath} 
\usepackage{amssymb} 
\usepackage{xcolor}
\usepackage{gensymb} 
\usepackage{tcolorbox} 

\newcommand{\refstyle}[1]{{\itshape\bfseries\color{eLifeMediumGrey}#1}}

\title{Cellular organization in lab-evolved and extant multicellular species obeys a maximum entropy law}

\author[1*]{Thomas C. Day}
\author[2]{Stephanie S. H{\"o}hn}
\author[1,3,4]{Seyed A. Zamani-Dahaj}
\author[1]{David Yanni}
\author[3]{Anthony Burnetti}
\author[3,5]{Jennifer Pentz}
\author[2\authfn{3}]{Aurelia R. Honerkamp-Smith}
\author[2\authfn{4}]{Hugo Wioland}
\author[2\authfn{5}]{Hannah R. Sleath}
\author[3*]{William C. Ratcliff}
\author[2*]{Raymond E. Goldstein}
\author[1*]{Peter J. Yunker}
\affil[1]{School of Physics, Georgia Institute of Technology, Atlanta, Georgia, United States}
\affil[2]{Department of Applied Mathematics and Theoretical Physics, Centre for Mathematical Sciences, University of Cambridge,
Cambridge, United Kingdom}
\affil[3]{School of Biological Sciences, Georgia Institute of Technology, Atlanta, Georgia, United States}
\affil[4]{Quantitative Biosciences Graduate Program, Georgia Institute of Technology, Atlanta, Georgia, United States}
\affil[5]{Department of Molecular Biology, Ume{\aa} University, Ume{\aa}, Sweden}

\corr{day.cooper.tom@gmail.com}{TCD}
\corr{sh753@cam.ac.uk}{SSH}
\corr{R.E.Goldstein@damtp.cam.ac.uk}{REG}
\corr{william.ratcliff@biology.gatech.edu}{WCF}
\corr{peter.yunker@gatech.edu}{PJY}

\presentadd[\authfn{3}]{Department of Physics, Lehigh University, Bethlehem, PA, United States}
\presentadd[\authfn{4}]{Institut Jacques Monod, Universit{\'e} de Paris Diderot/CNRS, Paris, France}
\presentadd[\authfn{5}]{Department of Chemistry, Imperial College London, London, United Kingdom}

\begin{document}

\maketitle

\begin{abstract}
The prevalence of multicellular organisms is due in part to their ability to form complex structures. How cells 
pack in these structures is a fundamental biophysical issue, underlying their functional properties. However, much 
remains unknown about how cell packing geometries arise, and how they are affected by random noise during growth - 
especially absent developmental programs. Here, we quantify the statistics of cellular neighborhoods of two different 
multicellular eukaryotes: lab-evolved ``snowflake'' yeast and the green alga \textit{Volvox carteri}. We find that 
despite large differences in cellular organization, the free space associated with individual cells in both organisms 
closely fits a modified gamma distribution, consistent with maximum entropy predictions originally developed for 
granular materials. This `entropic' cellular packing ensures a degree of predictability despite noise, facilitating 
parent-offspring fidelity even in the absence of developmental regulation. Together with simulations of diverse growth morphologies, these results suggest that 
gamma-distributed cell neighborhood sizes are a general feature of multicellularity, arising from conserved 
statistics of cellular packing.
\end{abstract}

\section{Introduction}
The evolution of multicellularity was transformative for life on Earth, occurring in at least 25 separate 
lineages \citep{Grosberg2007}. The success of multicellular organisms is due in part to their ability to assemble 
cells into complex, functional arrangements. Self-assembly, however, is fundamentally subject to random 
noise \citep{Zeravcic2014, Szavits-Nossan2014, Damavandi2019} that affects the final emergent structure \citep{Michel2019}. 
The physiology of multicellular organisms can depend sensitively on the geometry of cellular packing 
\citep{Bi2015a, Drescher2016, Jacobeen2018a, Larson2019, Schmideder2021}, and such noise may therefore have 
direct consequences on organismal fitness. Understanding the evolution of multicellularity, and the subsequent 
evolution of multicellular complexity \citep{BellMoers}, requires understanding the impact of random noise on 
multicellular self-assembly. How do organisms accurately assemble functional multicellular components in the 
presence of noise?

Recent work has shown that extant multicellular organisms can either suppress \citep{Hong2016} or leverage \citep{Haas2018} 
variability in the process of reliably generating structures, and their tissues can change function based on cellular packing 
geometry \citep{Bi2015}. This occurs through a coordinated developmental process involving genetic \citep{Davidson2001}, 
chemical \citep{Sampathkumar2020}, mechanical \citep{Deneke2018}, and bioelectric \citep{Levin2004} feedbacks between 
interacting cells. However, even with coordinated developmental processes, noise during self-assembly results in 
deviations from perfectly regular 
structures. Further, as these developmental processes have not yet evolved in nascent multicellular organisms, 
it is unclear how unregulated assembly can reliably result in reproducible packing geometries and multicellular structures.

Multicellular organisms also exhibit diverse growth morphologies; for example, cells can remain attached through 
incomplete cytokinesis \citep{Bonner1998, Grosberg2007, Knoll2011}, they can adhere through aggregative bonds \citep{Claessen2014}, 
and they can assemble multicellular groups through successive cell division within a confining membrane \citep{Angert2005, Herron2019}. These 
growth morphologies can have distinct intercellular connection topologies \citep{yanni2020topological}, changing how randomness 
is manifested. For instance, groups that grow with persistent mother-daughter bonds maintain the same intercellular 
connections, `freezing' in place any structural randomness that arises during reproduction. In contrast, cells in 
aggregates can rearrange, so their final structure emerges from a combination of reproduction and intercellular 
interactions and noise \citep{Delarue2016, Hartmann2019}. Further, the dimensionality of multicellular groups can 
vary, from quasi-two-dimensional sheets \citep{Brunet2019} to groups that grow equally in three 
dimensions \citep{Ratcliff2012, tang2020one, butterfield2000bangiomorpha}. While the impact of noise on systems 
in thermal equilibrium is well known to depend sensitively on spatial dimensionality 
\citep{Mermin1966, Hohenberg1967b, Vivek2017a}, no such information is yet at hand for biological development, 
which is intrinsically out of equilibrium. The growth morphology, connection topology, and dimensionality 
therefore altogether determine a multicellular architecture. Randomness resulting from many sources, such 
as stochastic cell division, variability in cell growth, intercellular interactions, and more, subsequently 
occurs as perturbations to this idealized 
form. It would appear that noise manifests in a unique, context-dependent manner in each of these different 
multicellular systems.

Here, we provide experimental evidence that, rather than being context-dependent, fluctuations in cell packing 
geometry instead follow a universal distribution, independent of the presence or absence of developmental regulation. 
We quantify the distributions of cellular space in two different types of organisms: experimentally-evolved 
multicellular yeast \citep{Ratcliff2012} and wild-type multicellular green algae \citep{GoldsteinARFM}. In 
both cases, maximum entropy considerations \citep{Aste2008} (see inset box) accurately predict the cell packing 
distribution. Building on these observations, we use computational models of diverse prescribed growth rules, 
mimicking extant biological morphologies, to show that cells are ubiquitously packed according to the maximum 
entropy principle. Detailed analysis of the case of green algae shows that correlations, \textit{i.e.}, the 
lack of structural randomness, produce  deviations from maximum entropy predictions, but that even a relatively 
small amount of randomness is sufficient to generate cellular packings that largely follow maximum entropy 
predictions. Next, we explore the evolutionary consequences of cell packing. We use the cell packing distribution 
to predict the distribution of snowflake yeast group sizes, an emergent multicellular trait that arises from cell 
crowding \citep{Jacobeen2018a}. Then, we use a theoretical analysis to show that the effects of fluctuations in 
intercellular space on the motility of green algae are small. 
These findings together suggest that, rather than impeding innovation, fluctuations in cell packing are highly 
repeatable, and may play a fundamental role in the origin and subsequent evolution of multicellular organisms.

\section{Results}

\begin{figure}
\begin{fullwidth}
\begin{center}
\includegraphics[width=0.91\linewidth]{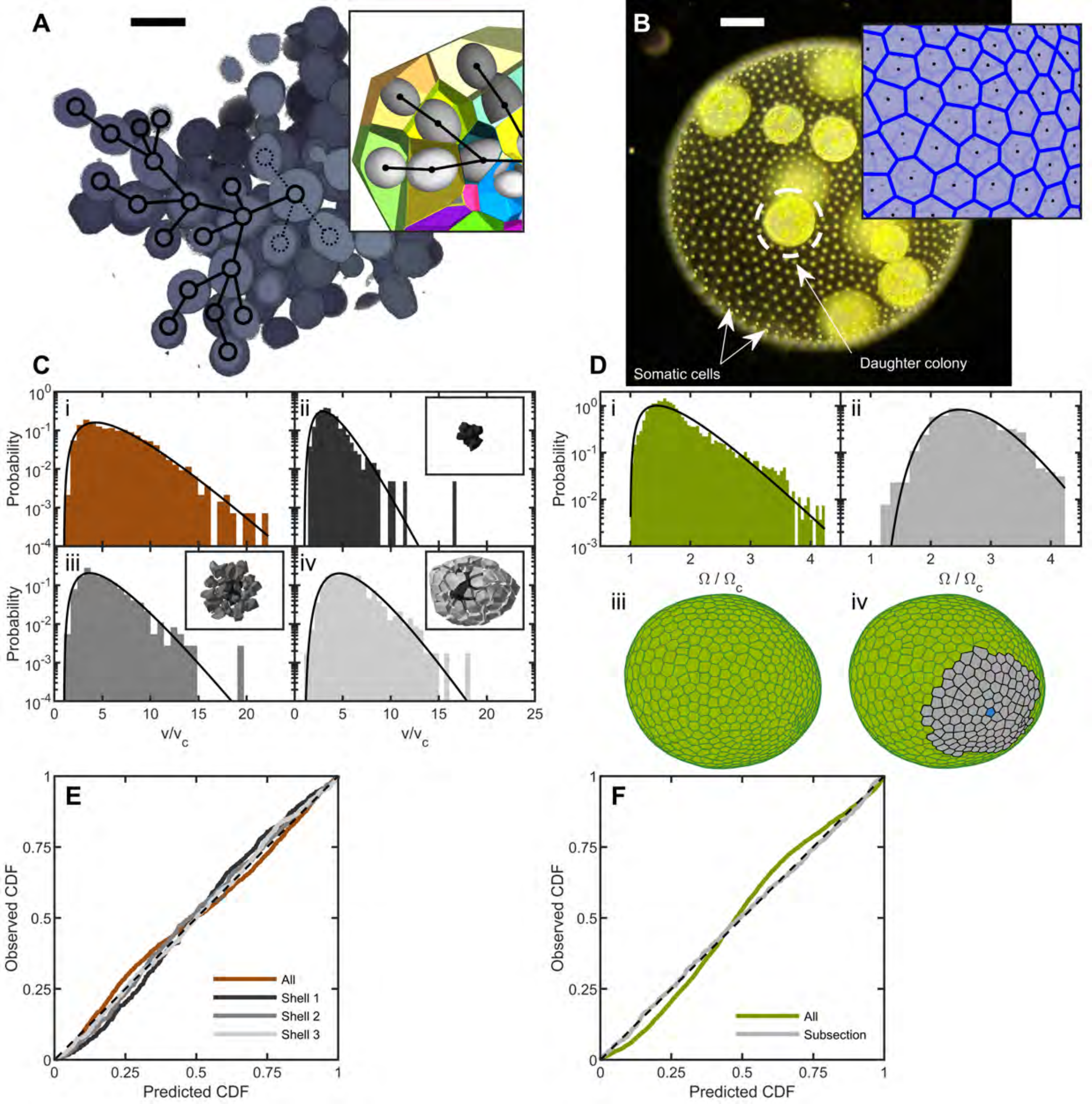}
\end{center}
\caption{Cell packing in two multicellular species. \textbf{(A)}, Cross section of a multicellular yeast organism, 
which grows with persistent intercellular bonds. Scalebar is $\SI{5}{\micro\meter}$. The inset shows a smaller 
section, with ellipsoidal fits to individual cells along with their corresponding Voronoi polyhedra. Black overlays 
indicate the connection topology between yeast cells; not all connections are labeled. \textbf{(B)}, Darkfield microscopy 
image of \textit{Volvox carteri}, scalebar is $\SI{100}{\micro\meter}$. Inset: a small piece of the Voronoi-tessellated 
surface; black points are somatic cell positions. \textbf{(C)}, Distributions of Voronoi polyhedron volumes 
as a function of cell size normalized by average size $v_c$ for snowflake yeast. In orange is the histogram for all cells; 
the other three distributions correspond to different subsections of Voronoi volumes. The cells were grouped 
into spherical shells with radius 
$R$ and width $\Delta R$ from the cluster center of mass. Shown are shells with edges $[0,6.2)$, $[6.2,9.7)$, and 
$[9.7,20.4)$ $\SI{}{\micro\meter}$. Black lines are maximum entropy predictions. \textbf{(D)}, Distributions of solid angles 
subtended by \textit{Volvox} somatic cells divided by a minimum solid angle $\Omega_c$. Solid black lines are the 
maximum entropy predictions. The top row shows the histogram for all cells in green and a subsection of correlated areas in gray. 
Bottom row illustrates the subsectioning process: blue polygon is the center of the subsectioned region. Only the Voronoi 
polygons, \textit{i.e.} not the somatic cells, are shown for clarity. \textbf{(E,F)}, Empirical cumulative distribution 
function vs entropic predictions for all distributions shown in \textbf{c,d}. The dashed black line represents hypothetical 
perfect agreement between observation and prediction.}
\label{fig:fig1}

\figsupp{Random cell budding positions in multicellular yeast groups. \textbf{(A)}, Bud scars determine the position 
of new cell buds, and are distributed across the surface of yeast cells. We locate bud scars in a spherical coordinate 
system with polar angle $\theta$ and azimuthal angle $\phi$. \textbf{(B)} Distribution of measured polar angle 
positions of new cells. \textbf{(C)} Distribution of measured azimuthal angle positions.}
{\includegraphics[]{Day_etal_figure1_supp1.pdf}}
\label{figsupp:fig1supp1}

\end{fullwidth}
\end{figure}

\begin{tcolorbox}
\textbf{Maximum entropy}

Within statistical physics, the maximum entropy principle relates randomness in low-level units (e.g., cells) to the 
properties of the assembly (e.g., a multicellular group). It works by enumerating all low-level configurations that 
conform to a set of constraints. Any particular group-level property can be generated by many different low-level 
configurations, but some group-level properties may correspond to more low-level configurations than others. Those 
that are generated by many configurations are more likely to be observed than those that correspond to relatively few 
configurations; in this way, the maximum entropy 
principle allows one to calculate the probability of observing different group properties, given a set of constraints. Multicellular groups obey 
a simple but universal constraint: each group has some total volume, $V$. 
This volume can be divided into $N$ pieces, where $N$ is the total number of cells. Each piece is associated with a 
particular cell, and the $N$ pieces must sum to the total volume of the group, $V = \sum_i v_i$, for $i=1,2,...,N$. 
Using this constraint, and assuming no correlations, one can predict the most likely distribution of volumes for the 
$N$ pieces. This approach has been successfully used to predict the distribution of free volumes within granular 
materials and foams \citep{Aste2008, Katgert2010}. Here we use it to predict the distribution of cellular free 
volumes in the absence of spatial correlations in cell positions.

\vspace{2mm}

Consider the ensemble of all possible cellular configurations in a simple group. As first derived 
by \citep{Aste2008} and \citep{Aste2007} for granular materials, the maximum entropy 
probability distribution $p(v)$ of cell neighborhood volumes within $V$ is the modified gamma distribution
\begin{equation}
	p(v) = \frac{k^k}{\Gamma(k)}\frac{(v - v_c)^{k-1}}{(\bar{v}-v_c)^k} 
	\exp\left(-k\frac{v-v_c}{\bar{v}-v_c} \right)
	\label{eq:kgamma}
\end{equation}
where $\bar{v}$ is the mean cell neighborhood volume, $v_c$ is the minimum cell neighborhood volume, $\Gamma(k)$ 
is the gamma function, and $k\equiv (\bar{v} - v_c)^2/\sigma_v^2$ is a shape parameter that is defined by $v_c$, 
$\bar{v}$, and the variance of the cell neighborhood volumes, $\sigma_v^2$. This distribution is expected if cell 
neighborhood volumes are determined independently of each other (while still conforming to the total volume constraint). 
In other words, volumes must be set randomly; correlations between the size of separate volumes will lead to deviations 
from maximum entropy predictions. If this condition holds, then maximum entropy volume distribution predictions should 
be valid, regardless of other geometric or structural details. For example, maximum entropy statistics hold in granular 
materials, despite the fact that they must obey strict force and torque balance conditions 
\citep{Aste2008, Snoeijer2004, Bi2015}. Further, the same approach applies to groups with a constraint 
on total area or length; this does not change the result, and $V$ can be replaced by $A$ or $L$ without other modifications.

\vspace{2mm}

In practice, we divide the total group volume or area into $N$ pieces via a Voronoi tessellation. The size 
of the space associated with cell $i$ includes the cell itself and the portion of intercellular space closer 
to its center than to the center of any other cell. As cells must have non-zero size, we therefore set $v_c$ to be 
the volume of a single cell without any intercellular space (or $a_c$, the area of a single cell).
\end{tcolorbox}

\subsection{Experimental tests of multicellular maximum entropy predictions}
To test whether different kinds of multicellular groups pack their cells according to the maximum entropy principle, 
we investigated cell packing in two different multicellular organisms. First, we used experimentally-evolved `snowflake’ 
yeast \citep{Ratcliff2012}, a model system of undifferentiated multicellularity. Second, we used the green microalga 
\textit{Volvox carteri}, a member of the volvocine algae that first evolved multicellularity in the Triassic
\citep{Starr1969,herron2009triassic}.

\subsubsection{Snowflake yeast}
Snowflake yeast grow via incomplete cytokinesis, generating branched structures in which mother-daughter cells remain 
attached by permanently bonded cell walls (\FIG{fig1}\refstyle{A}). New buds appear on ellipsoidal cells at a polar 
angle $\langle\theta\rangle=42\degree\pm 23\degree$ and azimuthal angle $\phi$ that is randomly distributed 
[$\langle\phi\rangle=180\degree\pm 104\degree$, \FIGSUPP[fig1]{fig1supp1}]. Therefore, cells bud in random orientations 
throughout the cluster. Due to the apparent absence of correlations, we expect that this structural randomness 
produces predictable distributions of cellular neighborhood volumes.

To determine the distribution of cell neighborhood volumes, we first must measure the position of every cell in a cluster. 
It is difficult to image individual cells within snowflake yeast clusters due to excessive light scattering. Instead, we used a 
serial block face scanning electron microscope equipped with a microtome to scan and shave thin ($\SI{50}{\nano\meter}$) 
layers off a resin block with embedded yeast clusters with stained cytoplasms. This process allowed us to determine the 
3D structure of snowflake yeast clusters and locate cell centers with nanometer precision. 

We define the group volume as the smallest convex hull 
that surrounds all cells in the cluster and computed the 3D Voronoi tessellation of cell centers within that 
(\FIG{fig1}\refstyle{A}). The distribution of cellular Voronoi volumes closely matched the predicted 
k-gamma distribution (\FIG{fig1}\refstyle{C}, $k = 2.88$). This agreement is quantified via ``P-P plots'' 
of the empirical cumulative distribution function (CDF) plotted against the predicted k-gamma CDF.  We find a 
root-mean-square residual $r_{RMS} = \sqrt{\langle(F(v)-F_i)^2\rangle} = 0.02$, where $F_i$ is the empirical CDF and $F(v)$ is 
the predicted k-gamma CDF.

The influence of the convex hull on these results was investigated by using an alternative procedure in which 
the Voronoi volumes were binned into shells centered at the cluster's center of mass (\FIG{fig1}\refstyle{C,E}). 
We binned cells into shells with shell edges of $[0,6.2)$, $[6.2,9.7)$, and 
$[9.7,20.4)$ $\SI{}{\micro\meter}$ away from the center of mass. We found that the distribution of Voronoi 
volumes within each shell matched the predicted k-gamma distribution, 
with $r_{RMS} = \{0.037, 0.020, 0.014\}$, $k=\{3.45, 3.08, 4.63\}$ in the shells shown in \FIG{fig1}\refstyle{C(ii-iv)}.

\subsubsection{Volvocine algae}
To test if cell neighborhood volumes in extant multicellular organisms are consistent with maximum entropy cell packing 
predictions, we examined cell packing within the green microalgae \textit{Volvox carteri}. Development in 
\textit{V. carteri}, which evolved over millions of years, is highly regulated, occurring through a stereotyped morphological 
progression \citep{Kirk2005}. \textit{V. carteri} embryos arise as a spherical cellular monolayer from palintomic cell 
divisions with incomplete cytokinesis, which leaves the cells attached via cytoplasmic bridges. These bridges disappear 
when ECM is secreted by the cells, filling the entire sphere, and eventually moving the cells apart. The approximately 
$1000$ somatic cells remain embedded on the surface of a translucent sphere of extracellular matrix (\FIG{fig1}\refstyle{B}). 
While six-fold coordination is the most frequent local arrangement of somatic cells, the fact 
that the cells are embedded in 
a surface with spherical topology requires there to be ``defects" with differing coordination number (e.g. $5,7$), and
these are found interspersed around the spheroid.  Thus, despite their developmental regulation, somatic cells exhibit 
a degree of disorder with respect to coordination number.  From a physics perspective, the local hexatic order
in the somatic cell arrangement is low (see Methods).

To determine the distribution of \textit{Volvox} cell neighborhood sizes, we imaged somatic cells using their chlorophyll 
autofluorescence in a light sheet microscope. Since the somatic cells are arranged around a surface embedded in 3D space, 
we constructed a 2D Voronoi tessellation of somatic cells on the surface. Each organism imaged had a different size, and 
therefore had a different mean Voronoi area $\langle A\rangle$. To compare distributions across organisms, we removed the 
systematic area differences by recording the solid angle $\Omega_i=4\pi A_i/S$ subtended by each somatic cell, where 
$S=\sum_i A_i$ is the total surface area of the organism. We found that the k-gamma distribution largely matched the 
distribution of solid angles (\FIG{fig1}\refstyle{D}, $k = 2.40$, $r_{RMS}=0.04$). However, there are 
systematic deviations between the data and maximum entropy predictions (\FIG{fig1}\refstyle{F}). 

We next investigated if maximum entropy predictions are more accurate within subregions with similar mean solid angles;
specifically we examine regions whose mean is
$\langle\Omega\rangle=0.0185\pm 0.0003$,
obtained across six organisms. The distribution of Voronoi solid angles within 
these subregions closely follows the k-gamma distribution (\FIG{fig1}\refstyle{F}, $k = 10.66$, $r_{RMS}=0.01$). 
This observation suggests that while there are systematically correlated subregions of cells, within these subregions 
cells are largely arranged randomly. Thus, the organization of \textit{Volvox carteri} somatic cells is consistent 
with maximum entropy predictions.

\subsection{Simulations of different growth morphologies}
\begin{figure}
\begin{fullwidth}
\includegraphics[]{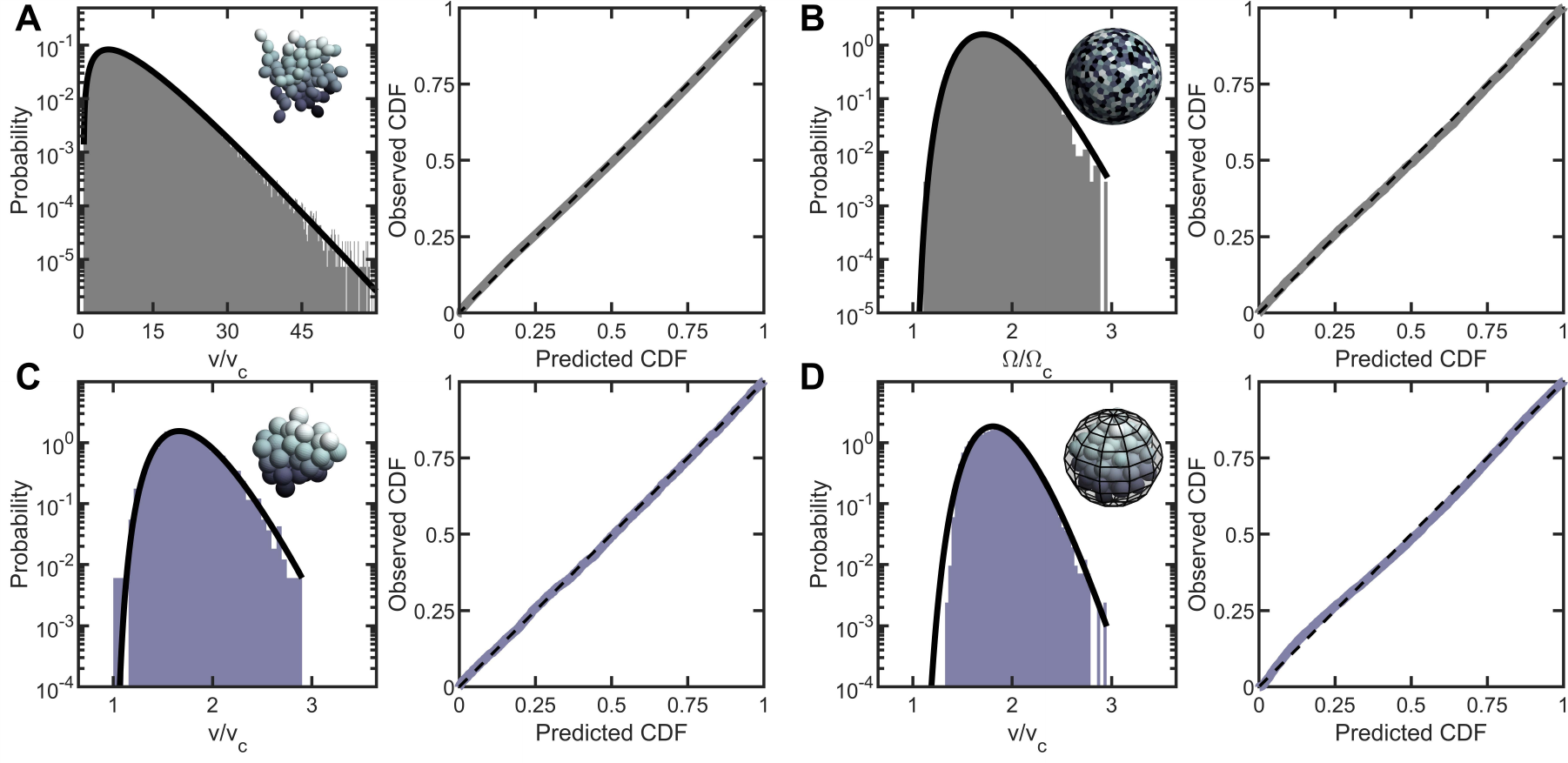}
\caption{Entropic packing is a general feature of simple multicellularity. We simulated four different growth morphologies: 
\textbf{(A)} Tree-like groups formed with rigid, permanent bonds between cells, \textbf{(B)} surface-attached cells located 
on a spherical surfac, \textbf{(C)} aggregates formed with attractive ``sticky'' interactions, and \textbf{(D)} groups formed 
by rapid cell division within a maternal membrane. In all subfigures, left panel shows the predicted and observed probability 
distributions, and right panel plots the observed cumulative distribution vs. the expected cumulative distribution. 
Histogram bars represent measured Voronoi volume distribution in simulations, and black solid line represents the maximum 
entropy prediction. Maximum entropy predictions accurately described the distribution of cellular volumes/areas, despite 
their varying mechanisms of group formation ($r_{RMS}\leq 0.01$).}
\label{fig:fig2}

\figsupp{Three different distributions were tested for goodness-of-fit: the maximum entropy prediction (black line), the 
normal distribution (red), and the log-normal distribution (blue).}
{\includegraphics[]{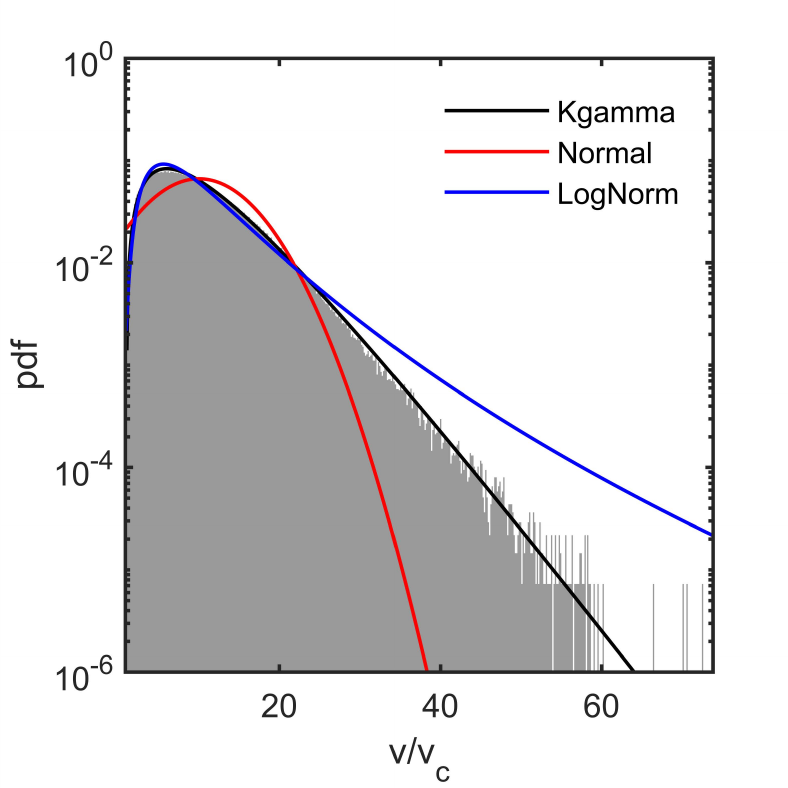}}
\label{figsupp:fig2supp1}

\end{fullwidth}
\end{figure}

We next used simulations to investigate the impact on cell packing of four different growth morphologies: growth via 
incomplete cell division (\textit{cf}. snowflake yeast), cells distributed on a spherical surface (\textit{cf. Volvox}), 
aggregation, and palintomy. The goal of these studies was to determine if morphological details and constraints impact 
entropic packing using simplified models that capture the essential features of the growth and behavior of these varied organisms.

These geometric simulations of multicellular groups that grow via incomplete cell division were inspired by previous simulations 
of snowflake yeast \citep{Jacobeen2018a, Jacobeen2018}. Daughter cells bud from mother cells with experimentally determined 
polar angle and random azimuthal angle, and remain attached to mother cells with rigid bonds. We ran $9,100$ simulations 
starting from a single cell, each of which underwent $7$ generations of 
division, and calculated the Voronoi tessellation of the final structure from each simulation. The distribution of Voronoi volumes closely 
matched the k-gamma distribution across four orders of magnitude (\FIG{fig2}\refstyle{A}, $k = 2.26$, $r_{RMS}= 0.007$). 

Inspired by {\it Volvox}, we simulated cells distributed across the surface of a sphere through a random Poisson point process. We completed 
$10$ simulations, each with $1000$ cells, and computed the distribution of solid angles subtended by Voronoi cells. As shown in 
\FIG{fig2}\refstyle{B}, the distribution of Voronoi 
solid angles is consistent with maximum entropy predictions ($k = 9.29$, $r_{RMS}=0.009$).

Next, we simulated organisms that stick together via reformable cell-cell adhesions, a mechanism of group formation that is common in 
biofilms and extant aggregative multicellular life \citep{Claessen2014} (\textit{i.e.}, \textit{Dictylostelium} and \textit{Myxococcus}; 
\FIG{fig2}\refstyle{C}). In these simulations, multicellular aggregates were grown from a single cell. Seven generations of cell division occured, 
in which new cells appear on the surface of existing cells at random positions, and steric interactions force cells to separate after division and 
occupy space.  Aggregative bonds were modeled through harmonic interactions of the cell centers. The observed Voronoi volume distributions were 
consistent with maximum entropy predictions ($k = 7.84$ and $r_{RMS}=0.007$).

Finally, we modeled cells undergoing palintomic division within a maternal cell wall, as is common in green algae 
\citep{Lurling1997, Boraas1998, Ratcliff2013, Fisher2016, Herron2019}, and is reminiscent of both baeocyte production in 
\textit{Stanieria} bacteria \citep{Angert2005} and neoproterozoic fossils of early multicellularity \citep{Xiao1998} (\FIG{fig2}\refstyle{D}). 
The details of these simulations remained similar to the simulations of aggregative multicellularity, with the important difference being that 
instead of harmonic interactions between cell centers enforcing groups to stay together, cells interacted with a spherical maternal wall 
acting as a corral. The Voronoi volume distributions for these simulations were also consistent with maximum entropy predictions 
($k = 15.16$ and $r_{RMS} = 0.013$).

Taken together, the results of these simulations suggest that a broad distribution of cell neighborhood sizes is a general feature of multicellular 
growth morphologies. In particular, when cell locations are random under these rules, cell neighborhood size distributions closely follow the 
k-gamma distribution.

\subsection{The role of spatial correlations}
While we have shown that the distribution of cell neighborhood volumes closely follows the k-gamma distribution in two very different 
organisms, we have also seen that in some cases maximum entropy predictions are more accurate in sub-sections of an organism than across 
its entirety. For instance, in \textit{Volvox} we observed that $r_{RMS}$ is much smaller within subregions with similar mean solid angles 
than across the whole organism. This observation suggests that correlations exist in the arrangement of \textit{V. carteri} somatic cells, causing 
deviations from maximum entropy predictions.

\begin{figure}[t]
\begin{fullwidth}
\includegraphics[]{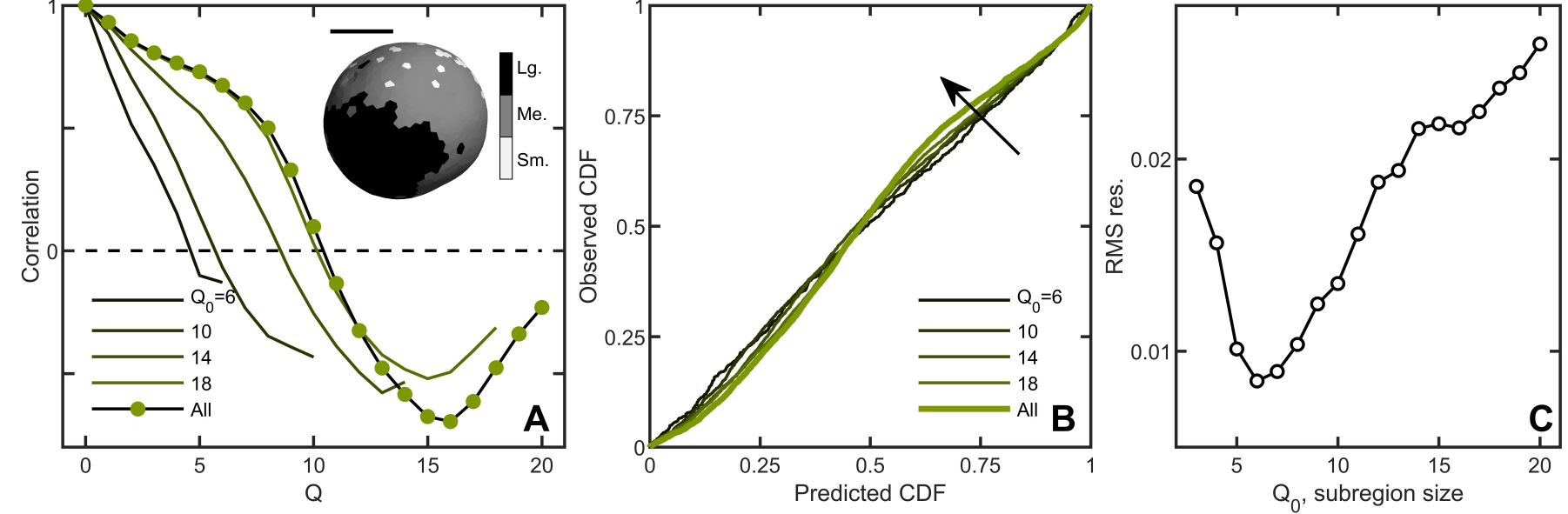}
\caption{Correlations lead to deviations from maximum entropy predictions in \textit{Volvox carteri}. \textbf{A} Correlation 
function of Voronoi polygon areas vs. network neighbor distance $Q$. Green circles represent all experimental {\it Volvox} data. 
Lines indicate the same correlation function calculated in subsections of size $Q_0=\{6,10,14,18\}$. Inset: visualization of spatial 
correlations of solid angle; one \textit{Volvox's} Voronoi tessellation is displayed with a three-color heatmap corresponding 
to polygons 
with areas smaller than (light gray), within (gray) and larger (black) than one standard deviation of the mean. Scale bar is 
$\SI{200}{\micro\meter}$. \textbf{B}, PP plots for the observed vs predicted cumulative distribution function. 
In green is the \textit{Volvox} 
distribution for all cells before corrections for correlations. A selection of differently-sized subsections is also plotted, corresponding to sizes 
$Q_0=\{6,10,14,18\}$. Arrow indicates direction of increasing $Q_0$ value. \textbf{C}, Root-mean-square residual deviation from maximum 
entropy predictions as a function of subsection size, as a function of nearest neighbor number $Q_0$. As the subsection size 
increases (including more and more uncorrelated Voronoi areas), the deviation from predictions first decreases 
until $Q_0=6$, then increases.}
\label{fig:fig3}
\end{fullwidth}
\end{figure}

The spatial correlations in the cellular areas in \textit{Volvox} were studied first by plotting a 3D heatmap of 
Voronoi solid angle sizes (\FIG{fig3}\refstyle{A}).  It is apparent that extended spatial regions have well-defined and non-random mean 
Voronoi solid angles. We quantified this feature by calculating the spatial correlation function $C(Q)$ of the solid angle 
\begin{equation}
    C(Q) = \frac{\langle \left(\Omega - \langle\Omega\rangle\right) Y_Q \rangle}{\sigma_\Omega \sigma_{Y_Q}},
\end{equation}
where $Y_Q=J(Q)^{-1}\sum_j (\Omega_j - \langle\Omega\rangle)$ is the average deviation of the solid angle of a given 
polygon's neighbors at a 
neighbor distance $Q$ from the mean.  Here, the number of neighbors is $J(Q)$, a function of $Q$, 
which enumerates the network distance from the polygon 
of interest (i.e. $Q=1$ calls the nearest neighbors, of which there are $J(1)$, $Q=2$ calls the next nearest neighbors, of 
which there are $J(2)$, and so on). The standard deviation of the solid angle across the population is $\sigma_\Omega$, and 
$\sigma_{Y_Q}$ is the standard deviation of $Y_Q$ across the population. We find that \textit{Volvox} Voronoi 
solid angles are positively 
correlated at distances as large as $Q=10$ (\FIG{fig3}\refstyle{A}). This analysis suggests that there are systematic differences 
in \textit{Volvox} group structure in different spatial regions. We therefore should expect to observe deviations 
from the k-gamma distribution, 
which was derived under the assumption that there are no correlations in the division of space among cells.

A natural question is whether maximum entropy predictions are more accurate within correlated subregions of an organism. 
We measured the Voronoi distribution in subregions with similar mean solid angles across six organisms and, for each 
subregion, a central node 
and its neighbors up to $Q_0$ were identified. We varied $Q_0$ from $3$ (corresponding to, on average, 38 cells in the 
subregion) to $Q_0=20$ (1016 cells on average in the subregion) to measure the Voronoi solid angle distributions in 
subregions of different sizes. 
Spatial correlations were weaker within smaller domains (\FIG{fig3}\refstyle{A}), and deviations from maximum entropy 
predictions smaller as well 
(\FIG{fig3}\refstyle{B}) with the minimum $r_{RMS}$ at $Q_0=6$ (\FIG{fig3}\refstyle{C}, average of 133 cells). These 
observations suggest 
that while there are systematic correlations between subregions, cell neighborhood sizes are largely randomly distributed 
within subregions.

\subsection{The crucial role of randomness}

How much randomness is necessary for the k-gamma distribution to predict cell neighborhood size distributions? Our analysis of the solid angle 
distribution of \textit{Volvox} cells demonstrates that maximum entropy principle predictions are relatively accurate ($r_{RMS}=0.04$) 
even in the presence of some spatial correlations. However, assessing the stability of entropic distributions to correlative perturbations is crucial 
to determine how broadly applicable entropic packing may be for multicellular organisms.
We investigated the stability of the maximum entropy distributions by simulating three different sources of correlations: 
(i) size polydispersity, (ii) defined growth patterns, and (iii) coordinated cellular apoptosis. In each scenario, we varied 
the relative strength of correlations and noise, and monitored how closely the cell neighborhood size distributions agreed with the 
k-gamma distribution via P-P plots and the $r_{RMS}$.

The impact of heritable size polydispersity was investigated by simulating aggregative groups consisting of large and small cells. 
All simulations were seeded with one small cell and one large cell. We then varied the probability $\xi$ that a new cell is the same size 
as its mother from $0.5$ to $1.0$. When $\xi=1$, cells always produce offspring with the same radius; for $\xi=0.5$ it is equally likely 
that a small cell produces a small or large daughter (and vice versa for large cells). Therefore, groups with $\xi=1$ have correlated regions of 
cell size, but the degree of correlations decreases with decreasing $\xi$. While groups with $\xi=1$ deviate significantly from the 
k-gamma distribution ($r_{RMS} = 0.09$), we observed that even a small amount of randomness results in excellent agreement between 
simulated groups and the k-gamma distribution (in order from $\xi = 1$ to $\xi = 0.5$, $r_{RMS} = \{0.09, 0.04, 0.03, 0.03, 0.02\}$).

\begin{figure}[t]
\begin{fullwidth}
\includegraphics[]{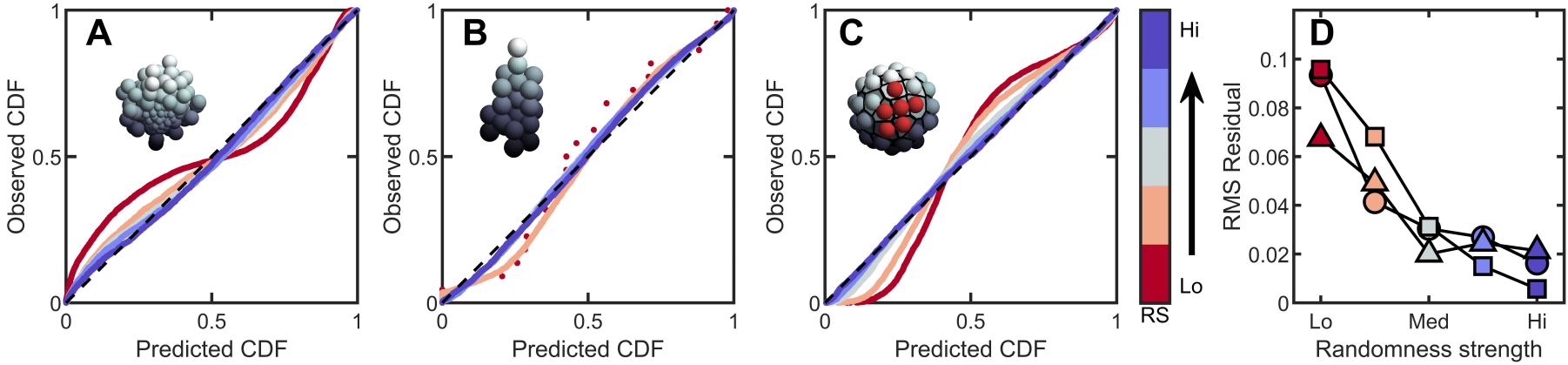}
\caption{Introducing correlations and structure can break the maximum entropy distribution. In \textbf{A-C} are PP plots of the observed 
vs. predicted cumulative distribution function for three different simulations. The colors correspond to increasing levels of noisiness in 
the simulations, from red (strongest correlations/determinism) to blue (strongest noise). The dashed black line in each represents $y=x$, 
or exact predictive efficacy. \textbf{(A)} Aggregative groups with bimodal size polydispersity; noise is introduced by varying the probability 
that small cells reproduce into small or large cells, and vice versa. \textbf{(B)}, Tree-like groups with persistent intercellular bonds 
that grow according to a growth plan modified by noise in cell placement. \textbf{(C)}, Surface-bound groups with programmed cell 
death events that may be localized or randomly dispersed. \textbf{(D)} The root mean square deviation from predicted values for each simulation case. 
Circles are aggregative simulations from \textbf{A}, triangles are tree-like simulations from \textbf{B}, and squares are surface-bound 
simulations from \textbf{C}.}
\label{fig:fig4}
\end{fullwidth}
\end{figure}

Next we investigated groups with varying amounts of noise on top of defined growth patterns. In these simulations, new cells bud in 
precise positions; the first daughter at the position $\theta=0\pm \eta$, $\phi=0\pm \eta$ in spherical coordinates), the second at 
$\theta=90\pm \eta$, $\phi=0\pm \eta$, the third at $\theta=90\pm \eta$, $\phi=180\pm \eta$, etc., where the noise is uniformly distributed 
with zero mean and width $\eta$. For $\eta=0$ (no noise), the distribution of Voronoi volumes was discontinuous, since cells could 
only access a finite number of local configurations. As expected, as $\eta$ increases ($\eta=\{0, 5, 30, 60, 90\}$), $r_{RMS}$ decreases
($r_{RMS}=\{0.07, 0.05, 0.02, 0.02, 0.02\}$).

Finally, we investigated groups with localized and random cell death. In these simulations, $50$ cells were confined to the surface of 
a sphere of unit radius following the protocol described above. One cell is randomly selected to die. Centered at this cell, a spherical 
region of radius $R$ is defined, and then $10$ cells in this region were randomly selected to die (and disappear, thereby not contributing 
to the Voronoi tessellation). For small $R$, cell death is highly localized, and thus spatially correlated. As $R$ increases, cell death 
events become less localized, and therefore more random. We find that highly correlated cell death resulted in large deviations from maximum 
entropy predictions. Conversely, as $R$ increases dead cells become less localized, the observed distribution becomes more accurately
described by the k-gamma distribution; as $R$ increases from $R=\{0.75, 1, 1.25, 1.5, 1.75,2\}$, we find $r_{RMS} = \{0.10, 0.07, 0.03, 0.02, 0.01\}$.

In summary, absent randomness, spatial correlations lead to large deviations from the k-gamma distribution. Yet, with even a small amount of 
randomness, the k-gamma distribution holds significant predictive power. These simulations suggest that maximum entropy predictions are 
likely to be robust against even moderate correlations.

\subsection{Parent-offspring fidelity via maximum entropy packing}
\begin{figure}
    \begin{fullwidth}
    \includegraphics[]{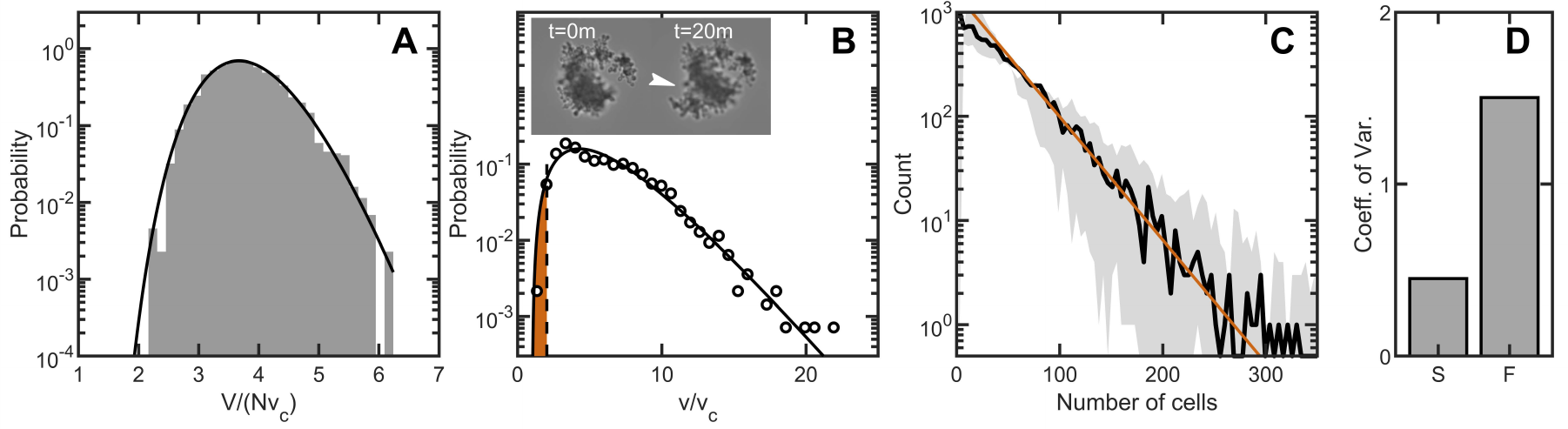}
    \caption{Maximum entropy cell packing generates a consistent and predictable life cycle in snowflake yeast. \textbf{(A)} 
    Distribution of total cluster volume for $3000$ simulated snowflake clusters, each with $N=100$ cells. The total volume is 
    divided by the minimum possible volume $Nv_c$. The k-gamma distribution (black line, $k = 23.0, r_{RMS} = 0.0043$) provides a good 
    description of the data. \textbf{(B)} Distribution of all experimental Voronoi volumes (black circles) and the maximum entropy 
    prediction (black line). The vertical black dashed line is the critical Voronoi volume $v^*=2.02$ predicted from simulations. 
    Orange filled region integrates up to $p^*$ the probability that any one cell occupies a volume less than $v^*$. 
    Insets: sequential brightfield microscope images of one yeast cluster undergoing group fragmentation. White arrowhead indicates 
    location of fracture point. The images measure \SI{150}{\micro\meter} across from top to bottom. \textbf{(C)}, Experimentally 
    measured yeast cluster size distribution (solid black line) along with the prediction from weakest link theory (orange line). 
    Gray region represents $1\sigma$ confidence bounds on the measured distribution from estimating the number of cells in a group. 
    \textbf{(D)}, Coefficient of variation in group radius ($\sigma/\bar{R}$) for snowflake groups (S) and flocculating groups (F). 
    Data from \citet{Pentz2020}.}
    \label{fig:fig5}
    \end{fullwidth}
\end{figure}

So far, we have shown that randomness in cellular packing leads to highly predictable packing statistics. Here we show that maximum entropy 
statistics can directly impact the emergence of a highly heritable multicellular trait, organism size.

Prior work has shown that the size of snowflake yeast at fragmentation is remarkably heritable - higher, in fact, than the traits of 
most clonally-reproducing animals \citep{Ratcliff2015}. The size to which snowflake yeast grow depends strongly on the aspect 
ratio of its constituent cells; more elongated cells allow the growth of larger clusters before strain from cellular packing causes group 
fragmentation \citep{Jacobeen2018a, Jacobeen2018}. Recently, experiments with engineered yeast showed that this emergent multicellular trait, 
group size, was in fact more heritable than the underlying cellular trait upon which it was based (cellular aspect ratio), despite the 
fact that the mutations engineered in this system only affected cellular aspect ratio directly \citep{Zamani-Dahaj2021.07.19.452990}. 
Simulations of multicellular chemotaxis observed a similar effect \citep{Colizzi2020}. While at first glance this may seem surprising, we show below 
that the high heritability of snowflake yeast group size arises from the direct dependence of size on the robust maximum entropy distribution 
of volume within groups.

Before addressing how fracture impacts the distribution of cluster sizes by impacting the number of cells within a group, we 
first must address fluctuations in size among clusters with the same number of cells. Given a number of cells $N$ in the cluster, 
variation in cell packing fraction results in variation of the total volume. The arguments given above for predicting the distribution 
of individual cell volumes also applies to the distribution of total volume \citep{Aste2008}; 
the distribution of total volume for clusters with the same number of cells should follow the k-gamma distribution. To generate enough 
clusters with identical $N$ to test this prediction, we used simulations. We generated $3000$ snowflake yeast clusters, each with
$100$ cells, and measured their total volumes. The distribution of volumes is consistent ($r_{RMS} = 0.0043$, $k = 23.0$) with the k-gamma distribution as 
shown in \FIG{fig5}\refstyle{A}. Further, these fluctuations in size are small compared to the differences in size gained via 
reproduction of cells or lost via fracture.


To predict the group size distribution, we consider the probability of fragmentation via a weakest-link model of fracture. As the 
location of new cells is random (see \FIGSUPP[fig1]{fig1supp1}), each new cell has a chance of causing intercellular bond fracture. 
It was previously observed that bonds only break if cells are highly confined, that is they have smaller Voronoi volumes; otherwise flexible 
cellular branches simply bend \citep{Jacobeen2018a}. We model fracture as occurring when a cell's Voronoi volume is below a critical value 
denoted by $v^*$ (\FIG{fig5}\refstyle{B}) such that its motion is completely restricted.  We measure $v^*$ from simulations that determine 
the maximum local packing density for groups with same cell size and shape distributions as seen in experiments (see Methods). The probability 
that a particular cell is confined to a Voronoi volume $v\le v^*$ is the integral
\begin{equation}
    p^* = \int_{v_c}^{v^*} p(v) dv.
\end{equation}
As each cell in a cluster of $N$ cells independently has probability $p^*$ of having $v\le v^*$ (and thus causing fracture), 
the probability of a cluster with $N$ cells not fragmenting is
\begin{equation}
P(N) = (1-p^*)^{N}
\label{eq:WeakLink}
\end{equation}
As we do not model the fate of products of fragmentation (\textit{i.e.}, the size of the separate pieces post-fracture), 
we expect the weakest link model to be more accurate for larger clusters than it is for smaller clusters. 

We measured group size for approximately $10,000$ snowflake clusters, all descendants of a single isolate, using a particle 
multisizer, and found strong agreement between the experimentally observed cluster size distribution and the weakest-link 
prediction (the coefficient of determination is $r^2=0.97$ for $\log(Counts)$ vs $N$) (\FIG{fig5}\refstyle{C}). Hence, the 
predictable statistics of entropic cell packing guides 
the distribution of group size among offspring of a single isolate.

For context, we compared the distribution of group size in snowflake yeast to that of flocculating yeast, which forms multicellular groups 
via aggregation. The multicellular size of flocculating yeast depends on the rate of collisions with other cells and groups of cells. 
The growth rate of aggregates is thus typically proportional to their size, as larger aggregates are more likely to contact 
more cells \citep{Pentz2020}. In fact, the maximum size of a flocculating yeast aggregate is bounded by the duration of aggregation, an 
extrinsic parameter, while the minimum size can be a single cell \citep{Stratford1992}. Using data from \citep{Pentz2020}, we compared 
the group size distributions of snowflake yeast and flocculating yeast grown in the same environmental conditions. We find that flocculating 
yeast groups exhibit a much larger coefficient of variation in size compared to snowflake yeast groups \citep{Pentz2020} (\FIG{fig5}\refstyle{D}). 
These results demonstrate that randomly assembled groups can exhibit more reproducible group traits than groups assembled with correlations.

\subsection{Multicellular motility is robust to cellular area heterogeneity}

One of the issues arising from the existence of the broad distribution of somatic cell areas in {\it Volvox} is the extent to which 
colony motility is affected by that heterogeneity. 
Each of the somatic cells at the surface of a {\it Volvox} colony has two flagella that beat at $\sim 30$ Hz, in planes that are 
primarily oriented in the anterior-posterior (AP) direction but with a slight lateral tilt that makes each colony spin around 
its AP axis. A longstanding focus in biological fluid mechanics of multicellular flagellates has been to understand the 
connection between the beating of the carpet of flagella that cover their surface and their self-propulsion. Measurements 
of the flow fields around micropipette-held \citep{Short} and freely-swimming colonies \citep{Drescher} have shown that 
despite the discreteness of the flagella, the flow is remarkably smooth, albeit often displaying metachronal waves \citep{Brumley2015}, 
long-wavelength phase modulations of the beating pattern.

A heuristic explanation for the smoothness of the flows can be developed by noting first that the flow arising from each flagellum, 
beating close to the no-slip surface of the colony, will fall off only as an inverse power of distance $r$ from the flagellum. Thus, 
the superposition of the flows from many flagella will be sensitive to contributions from distant neighbors and will tend to wash 
out local variations in flagellar actuation. This argument can be made quantitative using two different models for the motility of 
such flagellates. The first is the ``squirmer" model \citep{Lighthill}, in which the flagellate is characterized by a tangential 
``slip" velocity $u(\theta)$ on the surface, which can be thought of as corresponding to the mean motion of the flagella tips. 
Here, $\theta\in [0,\pi]$ is the polar angle with respect to the AP axis. In this approach the details of the fluid velocity 
profile below the tips are not resolved, and in particular the no-slip condition at the surface of the ECM is ignored.  In the 
second approach \citep{Ishikawa}, which builds on earlier work \citep{Short} that specified a force density at the colony surface 
instead of a slip velocity, there is a specified force density applied at some small distance above the no-slip colony surface, 
and the flow field below that locus is resolved.  This approach, termed the ``shear stress, no-slip" model, captures the very 
large viscous dissipation
that occurs in the region between the ECM and the locus of forcing. 
Within either of these two approaches above the effects of area inhomogeneities can be investigated by coarse-graining the 
flagella dynamics; either the local slip velocity $u(\theta)$ or the local tangential force density $f(\theta)$ has noise.

In the squirmer model, the swimming speed $U$ is \citep{Lauga} 
\begin{equation}
U=\frac{1}{2}\int_0^{\pi}\! d\theta \sin\theta\, u_{\theta}(\theta)V_1(\theta),
\end{equation}
where 
\begin{equation}
    V_n(\theta)=\frac{2}{n\left(n+1\right)}P_n^{'}\left(\cos\theta\right)\sin\theta,
\end{equation}
$P_n$ is the Legendre polynomial, and the prime indicates differentiation with respect to its argument. If we represent 
the effects of area inhomogeneities as noise in the slip velocity, then it is most natural to use $V_n$ as the basis 
functions for the tangential slip velocity, expressed as
\begin{equation}
    u_{\theta}(\theta)=\sum_{n=1}^{\infty}u_n V_n(\theta),
    \label{slipvelocity}
\end{equation}
where $V_n(0)=V_n(\pi)=0$, guaranteeing that the slip velocity vanishes at the anterior and posterior poles \citep{Short}. 
Accurate experimental measurements of the azimuthal velocity field of {\it Volvox} \citep{Fidelity} show that it is 
well-captured by that lowest mode, 
leading to a modest anterior-posterior asymmetry. From the orthogonality relation for the
$V_n$,
\begin{equation}
    \int\! d\theta \sin\theta V_1(\theta)V_n(\theta)=\frac{2n(n+1)}{2n+1} \delta_{1n},
\end{equation}
we see immediately that the contributions from all modes $n>1$ vanish identically, and thus the swimming speed is given 
identically by the amplitude of the lowest mode $V_1(\theta)=\sin\theta$, \begin{equation}
    U=\frac{2}{3}u_1.
\end{equation}
Thus, within the squirmer model, motility is essentially insensitive to area inhomogeneities. This result does not 
preclude effects of those higher modes, only that such effects will be on quantities other than the swimming speed, 
such as the nutrient uptake rate \citep{Magar}.

In the shear-stress, no-slip model, the velocity field in the region between the colony radius $R$ and the radius $R(1+\epsilon)$ 
at which the shear stress is applied is solved separately from that for $r>R(1+\epsilon)$ and the two flow fields are matched at 
$R(1+\epsilon)$ through boundary conditions of continuity in velocity and normal stress and the specified discontinuity in shear 
stress. Analogously to the expansion of the slip velocity in the squirmer model  \eqref{slipvelocity}, noise in that discontinuity 
can be expressed by assuming that the coarse-grained shear force applied by the flagella has spatial variations, and can be 
expanded in the form
\begin{equation}
    f_{\theta}(\theta)=\sum_{n=1}^{\infty}f_n V_n(\theta).
    \label{shearforce}
\end{equation}
The swimming speed again depends only on the lowest-order mode in this expansion, 
\begin{equation}
    U=\frac{2\epsilon R}{3\mu}f_1,
\end{equation}
and we again have insensitivity of $U$ to inhomogeneities in the area per somatic cell.

\section{Discussion}
In this paper, we demonstrated that universal cellular packing geometries are an inevitable consequence of noisy 
multicellular assembly. We measured the distribution of Voronoi polytope sizes in both nascent and extant 
multicellular organisms, and showed that they are consistent with the k-gamma distribution, which arises via 
maximum entropy considerations. Using simulations, we demonstrated that k-gamma distributions arise in many different 
growth morphologies, and do so requiring only a relatively small amount of structural randomness. Further, we showed 
that the distribution of cell neighborhood sizes can be used to distinguish the effects of randomness from the 
effects of developmental patterning. Finally, we demonstrated that consistent packing statistics can lead to highly 
reproducible, and thus heritable, multicellular traits, such as group size in snowflake yeast. Altogether, these 
results indicate that entropic cell packing is a general organizing feature of multicellularity, applying to 
multicellular organisms with varying growth morphologies, connection topologies, and dimensionalities.

The effect of random noise has been an important area of research in developmental biology 
\citep{Tsimring2014, Lander2011}. During development, cellular growth, reproduction, differentiation, and 
patterning combine to form a multicellular organism. Random noise introduced at any stage in this process can 
result in phenotypic variability, which may affect an organism's fitness \citep{Waddington2014}. But while some 
multicellular traits exhibit high variability, others are tightly conserved, leading to a wide body of research 
addressing the origin of mechanisms underlying robustness and stability, and the nature of feedback mechanisms 
that must be present to manage the large number of stochastic fluctuations in gene expression and 
growth \citep{Gregor2007, Haas2018, Hong2016, Sampathkumar2020, Deneke2018}. In this context, our results 
demonstrate that random noise can itself lead to highly reproducible multicellular traits such as the cell 
packing distribution.

Our observation that heritable properties can arise from random processes is reminiscent of the reproducible 
structures and phenomena generated by random noise in a wide range of physical \citep{Shinbrot2001, Manoharan2015} and 
biological systems \citep{Tsimring2014, Lander2011}. While it may be surprising that the distribution of free space in snowflake 
yeast and \textit{Volvox} follow the same k-gamma distributions despite the many differences between these organisms, 
this universality actually extends beyond multicellular organisms to non-living materials, such as those seen in granular 
materials and foams \citep{Katgert2010, Varadan2003, Aste2008}. This broad universality likely arises due to the simple 
requirements for application of the maximum entropy principle to packing; specifically, there must be a total 
volume, individual volumes cannot overlap, and volumes must be determined independently (subject to the total volume 
constraint). It is thus important to note that entropic packing is not necessarily adaptive; it can readily emerge 
as a consequence of random cellular reproduction or interactions. While entropic packing statistics may produce 
advantages in some cases, they could be neutral or detrimental in others.

An example of one possible advantage granted by entropic packing is the parent-offspring fidelity that arises from 
its ensemble statistics. Since both parents and their offspring are assembled through similar noisy processes, they 
achieve similar cell packing distributions. This statistical similarity therefore details at least one heritable 
multicellular trait that does not rely on genetically regulated multicellular development. Other multicellular traits 
that build on the cell packing distribution are similarly affected by this emergent process and could become heritable 
as well. Such parent-offspring heredity could play a crucial role in the evolutionary transition to multicellularity, 
providing a mechanism for nascent multicellular organisms to participate in the evolutionary process without first 
having to possess genetically regulated development. Over time, developmental innovation may arise via multicellular 
adaptation, modifying or replacing entropic cell packing as a mechanism of multicellular heredity. Consistent with 
this hypothesis, maximum entropy retains considerable predictive power in extant multicellular organisms such as 
\textit{Volvox}, animal embryos \citep{Alsous2018}, and epithelial tissue monolayers \citep{Atia2018}, each of which 
have canalized development. There may be other examples of highly-evolved organisms which pack cells according to maximum 
entropy predictions, and future work could address cell packing in, e.g., animal embryos, brain tissue, and more. 
Finally, as fragmentation is a common mode of multicellular reproduction 
\citep{Larson2019, Prakash2019, Angert2005, Keim2004, Koyama1977}, fracture driven by maximum entropy packing statistics 
may be relevant to organisms other than snowflake yeast.

The broad distributions in cellular volumes we have found in two very different types of organisms, with two very 
different modes of reproduction and growth, suggest that noise in developmental geometry may be an inevitable 
consequence of almost any microscopic mechanism.  In this sense, they may be just as unavoidable in 
biological contexts as thermal fluctuations are in systems that obey the rules of equilibrium statistical 
physics.  As an example, we recall the ``flicker phenomenon" of erythrocytes, in which the red blood cell 
membrane exhibits stochastic motions around its equilibrium biconcave discoid shape.
Thought for many years to be a consequence of specific biochemical processes associated with living systems, 
flickering was eventually shown by quantitative video microscopy \citep{flicker} to be consistent with equilibrium 
thermal fluctuations of elastic biomembranes immersed in water.  This was later confirmed by similar studies of shape 
fluctuations exhibited by large lipid vesicles \citep{SchneiderJenkinsWebb}. 
The generalization of these considerations to homeostatic 
tissues with cell division, rearrangements and apoptosis has also been considered \citep{Risler,Kalziqi2018}.
While such membrane systems may differ greatly in the specific values of their elastic modulus (and, indeed, of 
their microscopic membrane constituents), the viscosity of the surrounding fluid, and their physical size, 
the space-time correlation function of fluctuations about the equilibrium shape adopts a universal form in 
appropriately rescaled length and frequency variables.  

These results on equilibrium fluctuations provide a conceptual precedent for the results reported here. A 
central issue that then arises from our results is how to connect any given stochastic biochemical growth 
process defined at the microscopic level to the more macroscopic probability distribution function observed 
for cellular volumes. Mathematically this is the same question that arises in the theory of random walks, wherein 
a Langevin equation defined at the microscopic level leads, through suitable averaging, to a Fokker-Planck 
equation for the probability distribution function of displacements. Can the same procedure be implemented for growth laws?

\section{Acknowledgments}

Core Facilities at the Carl R. Woese Institute for Genomic Biology. W.C.R. was supported by 
NIH grant 1R35GM138030. This work was funded in whole, or in part, by the
Wellcome Trust (Grant 207510/Z/17/Z; REG \& SSH). For the purpose of open access, the authors have 
applied a CC BY public copyright license to any Author Accepted Manuscript version 
arising from this submission. This work was also supported in part by Established Career 
Fellowship EP/M017982/1 from the Engineering and Physical Sciences Research Council (REG).

\section{Methods}
\subsection{Yeast genotypes and growth morphology}

\subsubsection{Snowflake yeast genotypes}
Multicellular yeast groups were constructed from initially unicellular \textit{Saccharomyces cerevisiae}. Petite yeast 
groups (P-) were used in all experiments except those noted below. Snowflake yeast were engineered by replacing a 
functional copy of \textit{ace2} with a nonfunctional version as described in \citep{Ratcliff2015} (these modified 
genotypes will be referred to as either snowflakes or Ace2KO). Under daily selection for large size through settling 
in liquid media, groups can arise via a single mutation in the \textit{ace2} gene \citep{Ratcliff2012, Ratcliff2015}. 
When the \textit{ace2} gene is not expressed, the final stage of cell division is not completed, and mother-daughter 
cells remain attached at the chitinous bud site. Since all cells are attached directly to their mothers, snowflake 
groups form a fractal-like branched tree collective. To measure bud scar size, we used a unicellular strain of 
Y55 yeast; these measurements were only used to pick parameters for snowflake yeast simulations. 

\subsubsection{Yeast growth morphology}
\textit{S. cerevisiae} cells reproduce by budding, a type of asexual reproduction where a new cell extrudes from the 
surface of the parent cell. During budding, mother and daughter cells remain attached via a rigid chitinous bond; 
in unicellular yeast, chitinase will degrade this bond as the last step in cell division, releasing the daughter 
cell and leaving behind a ``bud scar’’ on the mother surface and a ``birth scar’’ defining the proximal hemisphere 
on the daughter’s cell surface. In all experiments, we use yeast expressing bipolar budding patterns \citep{Chant1995}. 
The bipolar budding pattern is characterized by bud sites that typically do not form along the equator of the cell. 
Usually, the first daughter buds near the distal pole. Subsequent budding sites are typically positioned along a 
budding ring defined by a polar angle $\theta$ (Figure 6). Some buds will ``backbud'' towards the mother cell (i.e. 
on the proximal end of the cell), but most buds are placed on the distal side. By contrast, the azimuthal positions 
of all buds appears to be randomly distributed.


\subsubsection{Growth conditions}
All experiments were performed on yeast grown for approximately 24 hrs in 10 mL of yeast 
peptone dextrose (YPD, 10 g/L yeast extract, 20 g/L peptone, and 20g/L dextrose) liquid medium at 30C, and shaken 
at 250rpm in a Symphony Incubating Orbital Shaker model 3500I. All cultures were therefore in the stationary phase 
of growth at the time of experiments.

\subsection{Scanning electron microscopy to measure group structure}
Since yeast cells have thick cell walls that limit the effectiveness of optical microscopy, we used a 
Zeiss Sigma VP 3View scanning electron microscope (SEM) equipped with a Gatan 3View SBF microtome installed 
inside a Gemini SEM column to obtain high resolution images of the internal structure of snowflake yeast groups and 
locate the positions of all cells. All SEM images were obtained in collaboration with the University of Illinois's 
Materials Research Laboratory at the Grainger College of Engineering. Snowflake yeast clusters were grown overnight 
in YPD media, then fixed, stained with osmium tetroxide, and embedded in resin in an eppendorf tube. A cube of 
resin $\SI{200}{\micro\meter}$ x $\SI{200}{\micro\meter}$ x $\SI{200}{\micro\meter}$ (with an isotropic distribution 
of yeast clusters) was cut out of the resin block for imaging. The top surface of the cube was scanned by the SEM to 
acquire an image with resolution $\SI{50}{\nano\meter}$ per pixel (4000 x 4000 pixels). Then, a microtome shaved 
a $\SI{50}{\nano\meter}$ thick layer from the top of the specimen, and the new top surface was scanned. This process 
was repeated until 4000 images were obtained so that the data cube had equal resolution in $x,y,z$ dimensions.

Custom image analysis scripts were written for the SEM datasets. First, a local adaptive threshold was used to 
binarize the image. A distance transform was used to identify the center of each cell slice in a particular 2d 
image. A watershed algorithm was then seeded with the cell slice centers, followed by a particle tracking algorithm 
to label cells across image slices. After labeling, the boundary for each cell was found, resulting in a point 
cloud of the exterior of each cell. Each cell was then fitted with an ellipsoid with nine fit parameters: 
$(x_0, y_0, z_0)$ cell center, $(a,b,c)$ cell radii, and $(\theta,\varphi,\psi)$ for cell orientation. The 
net rotation matrix $R$ was then found, where each column of $R$ corresponds to the direction vector of one principal 
axis of the ellipsoid. We consider the radii of the principal axes $(a,b,c)$ to be part of a diagonal scaling matrix 
$S$ which sets the ellipsoid size. Since the SEM images only capture the cell cytoplasm, each principal axis was increased 
in size by an additional $\SI{100}{\nano\meter}$ to account for the cell wall during visualization. Last, although there 
is no possible 3d 3x3 translation matrix, a 4x4 translation matrix $T$ can capture the position of the cell center 
$(x_0, y_0, z_0)$. Adding one additional column and row to the matrices $R$ and $S$ with the diagonal element being 
$1$ and all other elements being $0$ then means that a unit sphere centered at the origin can be mapped to any specific 
cell by a surface matrix $M = TRS$, and furthermore any point on the cell’s surface can be mapped back to the unit sphere 
by the inverse of $M$. Then, the surface matrices are the only information that must be stored. From this dataset, 
$20$ clusters of $105 \pm 51$ cells in each cluster were identified along with their intercellular mother-daughter chitin bonds.

\subsubsection{Petite yeast cell size and shape}
We measured cellular volumes from SEM images by ellipsoid fits. The average cellular volume of 
petite yeast was $v_c = \SI{17.44}{\micro\meter^3}\pm \SI{7.33}{\micro\meter^3}$. This measurement was 
used in our Voronoi distribution derivations. We measured the mean cellular aspect ratio to be $\alpha\equiv a/b = 1.28\pm 0.20$.

\subsubsection{Bud scar size}
We next measured the typical size of bud scars on the surface of Y55 yeast cells. Single cells were stained 
with calcafluor to highlight the chitinous bud scars (\FIGSUPP[fig1]{fig1supp1}). Confocal $z$-stacks were 
obtained on a Nikon A1R confocal microscope equipped with a 40$\times$ oil immersion objective. These images 
were visualized using the image processing software FIJI, and the 3d volume viewer plugin. To track the 
location and size of bud scars, a custom MatLab script was written to map the strongest calcafluor signals, 
since calcafluor makes bud scars brighter than other portions of the cell wall. Brightness isosurfaces then 
isolated the bud scars from the cell wall. Next, the isosurface points were rotated to the $x-y$ plane by 
finding its principal components in a principal component analysis. The rotated surface points were then fit with 
an ellipse, returning the major and minor axes. The average of the major and minor axes returned an average 
interior bud scar diameter of $\SI{1.2}{\micro\meter}$. This value was later used in simulations of yeast groups.

\subsubsection{Bud scar locations}
We measured bud scar positional distributions for petite yeast Ace2KO. Since the SEM does not image chitinous 
bud scars, we approximated bud scar positions as the closest point on a mother cell's surface to the corresponding 
daughter cell's proximal pole. We recorded $1990$ bud scar positions in polar coordinates, as defined in 
\FIGSUPP[fig1]{fig1supp1}. There is a clearly defined polar angle for the budding ring, while the azimuthal angle is 
uniformly distributed. The mean and standard deviations of the two angular coordinates were $\theta=42\degree\pm23\degree$, 
and $\varphi=180\degree\pm 104\degree$.

\subsection{Imaging \textit{Volvox}}
\subsubsection{Cultivation and Selective Plane Illumination Microscopy}
The \textit{V. carteri f. nagariensis} strain HK10 (UTEX 1885) was obtained from the Culture Collection of Algae 
at the University of Texas at Austin and cultured as previously described \citep{Brumley2014}. To visualise somatic 
cells, {\it V. carteri} spheroids were embedded in $1\%$ low-melting-point agarose, suspended in liquid medium and 
imaged using a custom-built Selective Plane Illumination Microscope \citep{Haas2018}. Each somatic cell is mostly 
filled with a single chloroplast. Chlorophyll autofluorescence was excited at $\lambda = \SI{561}{\nano\meter}$ and 
detected at $\lambda = \SI{570}{\nano\meter}$. To increase the accuracy with which we identify somatic cell positions, 
$z$-stacks of six spheroids were acquired from three different angles ($0$, $120$, $240$ degrees) and fused as 
described in the following paragraph.

\subsubsection{Registration of cell positions}
Positions of cells were registered based on fluorescence intensity using custom Matlab scripts. This was achieved 
by carrying out a 2D convolution of each frame of the $z$-stack with a basic kernel modelling the appearance 
of a cell - this was set to be an asymmetric double sigmoidal function. Cell segmentation was corrected 
manually. $Z$-stacks taken from different angles were roughly aligned using Fiji and the Matlab function 
fminsearch to minimise distances between the reproductive cells. This alignment was used as starting point 
for alignment of the somatic cells again using fminsearch. The positions of somatic cells were merged and averaged.

\subsection{Voronoi Tessellation}
We used a Voronoi tessellation algorithm to measure the distribution of cell neighborhood sizes in groups. We computed both 3D and 2D Voronoi tessellations.

\subsubsection{3D Voronoi Tessellations}
First, we computed 3D Voronoi tessellations within a defined boundary. These tessellations were performed for 
experimental snowflake yeast data from the SEM and simulations of 3D groups using the open-source Voronoi 
code Voro++ \citep{Rycroft2009}, wrapped in a custom MatLab script. Voro++ takes as input the Cartesian coordinates 
of the cell centers and the boundary of the shape within which to compute the tessellation. Without a boundary, 
all of the Voronoi cells located on the periphery would extend to infinity. We started the tessellation process 
by setting the input boundary to be a sphere; the Voronoi algorithm tessellated space within the spherical boundary. 
Then, pieces of the sphere were pared away until a Voronoi tessellation within the group's convex hull was obtained, as described in the next paragraph. 

The boundary sphere was centered on the cluster's center of mass. Its radius was the distance to the farthest 
cell center plus an additional $\SI{5}{\micro\meter}$. Upon tessellation within the sphere, each Voronoi polyhedron is defined by Cartesian vertices $\mathbf{r_j}$. We group these preliminary vertices by the cells 
to which they correspond, so that $Q_i = \{\mathbf{r_1}, \mathbf{r_2}, ...,\mathbf{r_m}\}$ is a list of the 
$m$ vertices corresponding to cell $i\in[1,N]$, $N$ being the total number of cells in the organism. We next 
computed the cluster's convex hull, which is the smallest convex polyhedron that contains all cell centers. We then 
extended the vertices of the convex hull by $\SI{3}{\micro\meter}$ outwards from the cluster center of mass so the boundary 
contained the entirety of each cell. This new boundary polyhedron, whose vertices are labeled $B$, defines the cluster 
boundary. We then found the intersection polyhedron, $Z_i = Q_i \cap B$ by taking the union of the dual of their 
vertices. This process thereby trims all Voronoi polyhedra to lie exclusively within the cluster's convex hull. 
The polyhedra $Z_i$ were the final Voronoi polyhedra used for the remaining data analysis.

\subsubsection{Voronoi tessellation on a sphere}
For Voronoi tessellations of cells on the surface of simulated spheres (see \FIG{fig3} and \FIG{fig4} of the main text), we used a 
built-in Matlab function called "voronoisphere" for Voronoi tessellations on a sphere.

\subsubsection{Voronoi Tessellation on Non-Spherical Surfaces}
We also computed 2D Voronoi tessellations on surfaces embedded in 3D space using custom-written 
MatLab functions. This approach was used for {\it Volvox} experimental data. Performing this computation 
with {\it Volvox} experiments presented a challenge as {\it Volvox} are roughly spherical, but with varying local curvature. 
It was therefore necessary to compute a Voronoi tessellation on an arbitrary surface.

The first step toward generating the proper Voronoi tessellation was computing the Delaunay triangulation 
of the cells on the surface (the Voronoi tessellation is the dual of the Delaunay triangulation). First, we found 
the Cartesian coordinates of each somatic cell (as described above), and normalized these coordinates so that all 
cell centers laid on the unit sphere. Then, a Delaunay triangulation of the normalized points was calculated. 
Edges of the triangulation that cut through the unit sphere were eliminated, and edges that laid along the 
sphere surface were kept. This Delaunay triangulation therefore mapped out the connectivity of the somatic 
cells. We then projected that triangulation onto the lumpy surface. The Voronoi polygon vertices are the 
circumcenters of each Delaunay triangle. Further, any edge shared between two Delaunay triangles denotes 
an edge shared between the Voronoi vertices associated with those two triangles. We found all edges connecting 
the Voronoi vertices. Next, connected edges were flattened so that each Voronoi cell was a 2D polygon. This 
step eliminates the curvature associated with the surface of the organism. However, we found that the distribution 
of Voronoi areas was unaffected by taking either the planar approximation or by approximating the area by
taking the local curvature into account -- the average difference between Voronoi areas when approximating the surface as a plane $A_p$ vs. approximating the surface as a spherical cap $A_s$ was found to be $\langle \frac{A_p-A_s}{A_p} \rangle = 0.001$, measured for one organism. Therefore, we used the flattened Voronoi polygons as the final tessellation shapes.

\subsection{Data analysis of Voronoi measurements}
In all cases, the output of the Voronoi algorithm is a list of Voronoi polytope sizes: in 3D, the measurements were the final 
Voronoi polyhedron volumes, while in 2D the measurements were polygon areas. Histograms of these sizes were generated 
to compare with the k-gamma distribution. As we observe cells in direct contact with each other, the minimum size 
of a Voronoi volume or area was defined by single cell measurements. For petite yeast cells, the mean cell size was calculated from the ellipsoid fits described above to be $v_c = \SI{17.44}{\micro\meter}^3\pm \SI{7.33}{\micro\meter^3}$. 
In simulations, the minimum volume was set by the defined cell radius; in bidisperse simulations, the minimum size was 
set by the volume of the smallest cells. 

We then calculated the expected maximum entropy distribution using only the 
mean and variance of the observed Voronoi volumes, $\bar{v}$ and $\sigma^2$, as inputs. Together with the minimum volume 
$v_c$, these measurements define $k=(\bar{v}-v_c)^2/\sigma^2$, a dimensionless shape parameter \citep{Aste2008}. The 
maximum entropy distribution was therefore not fit to the data using, for example, a least squares method, but inferred 
from the first two moments of the distribution.

\subsubsection{\it Volvox}
Along the surface of the \textit{Volvox} organisms, there are gaps between some of the somatic cells due to the Gonidia 
that lie beneath, but near the surface of the organism. These Gonidia effectively occupy space on the surface, making 
it inaccessible to somatic cells. We excluded all Voronoi cells that intersected these Gonidia gaps. We identified 
gaps in the soma cells by flagging Delaunay triangles with exceptionally high aspect ratios. Any Voronoi polygons that 
intersect the flagged Delaunay triangles were then flagged and later excluded from the dataset. The polygons were 
generally spatially clustered, indicating that the Gonidial gaps were being correctly isolated. Roughly 
$90$ polygons were excluded from each organism.

In \textit{Volvox} organisms, each cell is surrounded by extracellular matrix, so cells do not contact each other. 
Furthermore, each of the six organisms studied varied in diameter (standard deviation in diameter was 
$\SI{28.2}{\micro\meter}$), yet all contained roughly the same number of somatic cells, leading to systematic 
differences in average surface area per cell across the organisms. Quantitatively, the coefficient of variation 
of the diameter of the groups was $CV_D = 0.05$, while the coefficient of variation in the number of cells in 
each group was roughly $10$ times smaller, $CV_N = 0.006$. To counter the systematic size differences between organisms, 
we converted the Voronoi polygon areas into solid angles by dividing by the total surface area of each organism, 
$\Omega_i = A_i/S$; we then grouped all six organisms together into one histogram. We allowed the minimum solid angle, 
used in the k-gamma equation, to be a fit parameter in a least squares minimization procedure. There was one 
outlier cell with solid angle $\Omega = 0.0048$ steradians; the next two smallest cells had solid angles 
$0.0068$ and $0.0069$ steradians. We removed the outlier; the least squares minimization procedure then fit a minimum solid angle 
$\Omega_c = 0.0070$ steradians. We used this value for all further calculations. Just as in the 3D case, the 
mean and variance of the solid angle were measured to set the expected maximum entropy distribution.

\subsection{Cluster size distribution measurements}
Cluster sizes were measured using a Beckman Coulter Multisizer 4e particle analyzer in the Cellular Analysis 
and Cytometry Core of the Shared User Management System located at the Georgia Institute of 
Technology. Petite Ace2KO clusters were taken from steady state concentration in YPD and then submerged 
in electrolytic fluid and passed through a $\SI{100}{\micro\meter}$ aperture tube. The volume measured on 
the multisizer corresponds to the volume of electrolyte displaced by the cluster. The number of cells in 
each cluster was then estimated by $N = V/v_c$, where $V$ is the volume of organism measured by the 
Coulter Counter, and $v_c$ is the average cell volume from SEM measurements, $v_c = \SI{17.44}{\micro\meter}^3$.

\subsection{Cumulative Distribution Function statistics}
To quantify goodness-of-fit for predicted maximum entropy distributions, we compared the predicated 
cumulative distribution function (CDF), $F(x)$, to the empirical CDF, $F_i$, using P-P plots. 
Exactly predicted points will lie on the line $y=x$ in these plots. We measured the root-mean-square 
residual from the line $y=x$,
\begin{equation}
    r_{RMS}=\sqrt{\langle (F_i - F(x))^2 \rangle}
\end{equation}

\subsection{Measurements of $\Psi_6$ in {\it V. carteri}}
From the light sheet images of \textit{Volvox}, we obtained the Cartesian coordinates of each somatic cell. 
From Delaunay triangulation, we then obtained a list of every cell's closest neighbors. Each cell and its 
$NN$ nearest neighbors did not generally lie in a plane due to local curvature of the {\it Volvox} surface. 
We therefore calculated in-plane and out-of-plane components using principal component analysis. 
The in-plane components were then used to write the positions of each nearest neighbor in polar coordinates. 
The formula for calculating $\Psi_6$ is
\begin{equation}
    \Psi_6 = |\langle \frac{1}{NN}\sum_{j=1}^{NN}e^{6i\theta_j} \rangle|
\end{equation}
where $\theta_j$ defines the polar angle coordinates around the cell of interest and $\langle...\rangle$ denotes 
averaging over all cells. We calculated $\Psi_6$ separately for each of six different organisms; we report $\Psi_6 = 0.03\pm 0.01$.

\subsection{Correlation of Voronoi Areas}
In {\it Volvox} organisms, we calculated the spatial correlation of polygon areas. First, we extracted the list of 
cell neighbors from the Delaunay triangulation of the organism surface. Nearest neighbors were designated as 
living a network distance of $1$ away from a cell of interest; next nearest neighbors live a network distance of 
$2$ away from the cell of interest, etc. The number of neighbors a network distance of $Q$ away is then $J(Q)$, which is empirically 
determined. The network correlation function is then
\begin{equation}
    C(Q) = \frac{\langle (\Omega - \langle\Omega\rangle) Y_Q \rangle}{\sigma_\Omega \sigma_{Y_Q}}
\end{equation}
where $Y_Q=J(Q)^{-1}\sum_j (\Omega_j - \langle\Omega\rangle)$ is the average deviation of the solid angle of a given polygon's 
neighbors from the mean. The standard deviation of the solid angle across the population is $\sigma_\Omega$, and $\sigma_{Y_Q}$ 
is the standard deviation of $Y_Q$ across the population.

\subsection{Simulation methods}

\subsubsection{Simulations of snowflake yeast groups}
Simulations of snowflake yeast groups were adapted from previously published work by 
\citet{Jacobeen2018, Jacobeen2018a} that found simulations of snowflake yeast growth morphology 
accurately replicated experimentally measured cellular packing fractions and average group sizes. In the present work, 
cells were modeled as prolate ellipsoids of revolution with a semi-major axis $a = \SI{2.88}{\micro\meter}$ and 
semi-minor axis $b = \SI{2.29}{\micro\meter}$, characterized by the aspect ratio $\alpha\equiv a/b = 1.26$. Each 
generation, every cell attempted to reproduce; however, if new cells closely overlapped with existing cells 
(\textit{i.e.} their bud scars are closer than $\SI{1.2}{\micro\meter}$), they were eliminated. Setting the number 
of generations (for example, $7$) sets the maximum possible number of cells in the group at the end of the 
simulation ($2^7=128$), and roughly sets the expected number of cells in the group ($\sim 100$). In our 
simulations, cells were $80\%$ likely to bud first from the distal pole (i.e. $\theta = 0 \pm 10$ degrees). Subsequent 
cells budded at a polar angle $\theta$, and with an azimuthal angle randomly chosen from a uniform distribution $
\varphi\in[0,2\pi]$; in other words, after the first bud, cells generally appeared along a ``budding ring’’. There 
was a $20\%$ chance that the first bud would appear along this budding ring instead of exactly at the pole. After 
$3$ bud scars, there was a $50\%$ chance that new cells bud on the proximal side ($\pi-\theta$) instead of the distal side. 
The orientation of the new cell is determined by the surface normal to the mother cell at the position of the bud site; 
the major axis of the new cell lies along the surface normal.

To compare exhaustively the distribution of Voronoi volumes between simulations and the k-gamma distribution, 
we simulated $9,100$ clusters. In each simulation, clusters were allowed to grow for $7$ generations of cell division, 
corresponding to an average of $94.2\pm 10.9$ cells per cluster. The budding ring was defined by the polar angle 
$\theta=45\degree$, a close approximation to the experimentally measured mean polar angle. These simulations did 
not include intercellular forces. The cell centers were recorded and then Voronoi tessellations were made within 
each cluster's convex hull.

\subsubsection{Simulations of \textit{V. carteri}}
We simulated a {\it Volvox}-like group with $N=1000$ cells confined to the surface of a sphere. Cells were placed 
on the surface of a sphere of unit radius by randomly selecting polar and azimuthal coordinates in a Poisson point process. The process proceeds as follows: each new cell was randomly placed, and its distance from all other cells was calculated. If the new cell is within a threshold distance $d$ from any existing cell, it was removed and a new cell was placed elsewhere 
on the spherical surface. This process was iterated until all $1000$ cells were placed. We chose a minimum separation distance of $d=0.088$, which allowed reasonably 
rapid convergence. We then calculated the Voronoi tessellation and the correlation function as described above.

\subsection{Simulations of two additional growth morphologies}
We next sought to model two additional classes of growth morphologies: sticky aggregates and cells contained within 
a maternal membrane. In both simulations, cells were modeled as spheres with unit radius.

\subsubsection{Aggregative groups}
First, we considered a multicellular model of sticky aggregates, mimicking group formation in, for example, 
flocculating yeast and bacterial aggregates. In our simulations, groups were grown from a single cell. New spherical 
daughters appeared at a polar angle $\theta$ and azimuthal angle $\varphi$. Within each step, 
there was stochasticity in the budding location: cells would appear at $\theta = \theta_0 \pm 15\degree$. The azimuthal 
angle was always drawn from a uniform distribution on the interval $[0\degree,360\degree]$.

Cells interacted with both steric and attractive interactions in overdamped dynamics. Steric interactions 
were modeled through a harmonic potential when two cells overlapped, with a cutoff once cells were no longer overlapping.  That is, for two cells $i$ and $j$ (radii $R_i$ and $R_j$) separated by the vector
$\mathbf{r}_{ij}=\mathbf{r}_j-\mathbf{r}_i$, the steric force acting on cell $i$ from cell $j$ is
\begin{equation}
    \mathbf{F}_{ij} = 
    \begin{cases} 
      0 & |\mathbf{r}_{ij}| > (R_i+R_j)\\
      \kappa_{s} \left(\vert \mathbf{r}_{ij}\vert - (R_i+R_j)\right)\hat{\bf r}_{ij} & |\mathbf{r}_{ij}| \leq (R_i+R_j)
   \end{cases}
\end{equation}
Attractive interactions (i.e., sticky, aggregative bonds) were also modeled through a harmonic potential, but 
these interactions had both a lower bound and upper bound cutoff.
\begin{equation}
    \mathbf{G}_{ij} = 
    \begin{cases}
      0 & |\mathbf{r}_{ij}| > 2(R_i+R_j)\\
      -\kappa_{a} (|\mathbf{r}_{ij}|- a(R_i+R_j))\hat{\bf r}_{ij} & (R_i+R_j)\leq |\mathbf{r}_{ij}| \leq 2(R_i+R_j) \\
      0 & |\mathbf{r}_{ij}| < (R_i+R_j),
    \end{cases}
\end{equation}
where $a$ sets the location of the attractive well minimum. We used $a=0.9$, so that the attractive interactions 
allow a small amount of cell overlap.

\subsubsection{Size polydispersity}
In simulations in which we introduced size polydispersity, cells were allowed to reproduce into two separate 
sizes, $R_1 = 1$ and $R_2 = 2$. The probability of budding cells of the same size as the mother cell is 
denoted $\xi$. When $\xi=1$, the mother cell always produces cells of the same size, while when $\xi=0.5$, 
there is a $50\%$ chance that the mother cell produces a cell of size $R_1$ or $R_2$, independent of the radius of the mother. Simulations were seeded with a pair of contacting cells, one each of the two radii.
The simulation then proceeded with subsequent rounds of cell division and mechanical relaxation.

\subsubsection{Groups confined within a membrane}
In another common mode of group formation, cells divide repeatedly within a confining membrane. This 
type of group formation has been observed in experimentally-evolved multicellular algae derived from 
unicellular \textit{Chlamydomonas reinhardtii} \citep{Herron2019}, and is reminiscent of both baeocyte 
production in \textit{Stanieria} bacteria \citep{Angert2005}, and neoproterozoic embryo fossils \citep{Xiao1998}. 
In a simulation model, we adopted the essential components of this class of growth: groups grow from a single 
spherical cell, cells divide stochastically, and cells interact sterically with both a maternal cell wall and 
each other. Typically, palintomic cell division occurs rapidly, meaning that the packing fraction remains the 
same within the maternal cell wall. We simulated this by increasing the radius of the cell membrane after each 
cell division, but before allowing any mechanical relaxations.

Steric forces between a cell and the maternal cell wall were modeled as being proportional to the non-overlapping volume of the cell and the maternal cell wall. In other words, if a cell is not contacting the membrane, there is no force acting on it. However, if the cell is contacting the membrane, the force is proportional to how much of the cell volume lies outside the membrane. Each cell was assigned volume $v_c = 4/3*pi$. The overlapping volume of the cell and the membrane is labeled $v_i$. The force the cell experiences from the membrane is then
\begin{equation}
    \mathbf{F}_i = \kappa_m (v_c - v_i) \mathbf{\hat{r}}_i,
\end{equation}
where $\mathbf{\hat{r}}_i$ is a unit vector pointing to the center of the maternal membrane. Additionally, steric interactions between cells were calculated as described above for aggregative groups.

\subsubsection{Groups confined to a spherical surface}
Some groups form by arranging cells around a central core of extracellular matrix (ECM). To simulate such groups, 
we modeled a sphere of ECM with cells arranged randomly along the surface. Cell positions were chosen by 
selecting a position in spherical coordinates from uniform polar $\theta\in[0,\pi]$ and uniform azimuthal 
$\varphi\in[0,2\pi]$ distributions. The only rule implemented in cell placement is that no two cells can be located 
closer than two cell radii from one another. If a new cell is chosen to be located too close to any existing 
cells, it is eliminated and a new position is chosen. We iterated this process until $N$ cells were placed on the ECM surface.

First, we chose to place $N=50$ cells on the surface. Therefore, the maximum cell radius allowing all $50$ cells 
to be placed is $0.283$ units (where the total sphere has unit radius). We chose the cell radius to be $0.1980$ 
units, which allowed for reasonably rapid random placement of all $50$ cells (other choices of cell radii demonstrate 
qualitatively similar results). We then used a built-in Matlab spherical Voronoi tessellation algorithm to 
calculate the solid angle subtended by each cell.

\subsubsection{Simulated cellular apoptosis}
In simulations with apoptosis events, cell death occurred after group generation (as described in the above subsection on aggregative groups). Briefly, groups were generated by iterated generations of cell division starting from a single cell. After this process, one cell was chosen at random to die. 
Then, all cells within a localization radius $R$ were flagged. Of the flagged cells, $9$ more were chosen at
random to die. Therefore, small localization radii correspond to highly localized death events, where $10$ juxtaposed 
cells may die together. As the localization radius increases, there are more flagged cells, and therefore more randomness 
in cell death. All other cells were unaffected by the cell death process.

\subsubsection{Tree-like groups with precisely defined cell placement/location}
We also investigated groups with precisely defined growth patterns. The spherical cells were held together 
with fixed, chitin-like bonds. The first cell was placed at the origin. It then proceeded to bud $3$ daughter 
cells, each of which also budded subsequent cells. The exact budding pattern is described below.

Daughter cells were placed as follows. In spherical coordinates on the surface of the mother cell, 
the first daughter cell was placed at $(\theta=0\pm\eta$, $\phi=0\pm\eta)$, the second at $(\theta=90\pm\eta$, $\phi=90\pm\eta)$, 
and the third at $(\theta=90\pm\eta$, $\phi=270\pm\eta)$, where $\eta$ is the strength of random noise added.

The first daughter cell's coordinate system was rotated $90\degree\pm\eta$ around the $z$-axis from the mother cell; 
in other words, for the first daughter cell, $x\xrightarrow{}x'$, $y\xrightarrow{}y'$, and $z\xrightarrow{}z'$, 
where $\mathbf{x'}=\mathbf{R}_z(\pi/2+\eta)\mathbf{x}$, $\mathbf{y'}=\mathbf{R}_z(\pi/2+\eta)\mathbf{y}$, 
and $z'=\mathbf{R}_z(\pi/2+\eta)\mathbf{z}$, and $\mathbf{R}_z$ is the rotation matrix around the $z$-axis. 
This daughter cell then proceeded to bud daughters in the exact same pattern as its mother; however, because 
its local coordinates were rotated, the budding positions were also rotated $90\degree$ with respect to the mother 
cell's buds. This process was iterated for $5$ generations of cell division. When $\eta=0$, this corresponds to 
only $3$ cells overlapping $3$ other cells. The $3$ overlapping cells were then removed.

After each round of cell division, cells were allowed to relax mechanically in overdamped dynamics according to steric 
repulsive interactions and sticky, rigid bond interactions to their mother cell. The steric interactions were the 
same as described above. Fixed bond interactions were modeled as follows. When new cells appear, they incur a bud 
scar on the mother cell's surface and a birth scar on the daughter cell's surface. The positions
$\mathbf{r}_{bu}$ and $\mathbf{r}_{bi}$ of the bud scar and the birth scar were recorded and tracked. The vector pointing from the 
bud scar (on the mother's surface) to the birth scar (on the daughter's surface) is 
$\mathbf{r} = \mathbf{r}_{bi} - \mathbf{r}_{bu}$. Then, the force acting on a cell from it's mother cell is 
\begin{equation}
    \mathbf{F}_{mother} = \kappa \left(|\mathbf{r}|-2\right)\mathbf{\hat{r}}
\end{equation}
where $\kappa$ was the chitin bond strength. In addition, cells experienced forces from all of their daughter buds 
(given by the same relationship and the same chitin bond strength). The initially seeded cell did not experience 
forces from a mother cell.

For $\eta=0$ (i.e., no noise), the distribution of Voronoi volumes was visually discontinuous, since cells could 
only access a finite number of local configurations. As the noise strength increased, the maximum entropy predictions 
were gradually recovered.

\section{Appendix}

It may appear surprising that the distribution of cell volumes is not governed by the Central Limit Theorem (CLT), i.e. 
the volumes are not distributed normally. After all, Voronoi polytope volumes are generated from many randomly 
interacting pieces - should not these many different random fluctuations sum to a CLT-like scenario? A simple 
comparison between the modified gamma distribution, a normal distribution, and a log-normal distribution shows 
in fact that both the normal distribution and the log-normal distribution fail to capture essential characteristics 
of the volume packing, while the k-gamma distribution does. For snowflake yeast, the reason for this disagreement 
is that as each new cell is added to a cluster, it changes the entire volume distribution, since the new cell 
occupies space which was previously unoccupied. It therefore changes the volumes of all its nearest neighbors; 
if they flex to accommodate the new cell, then those neighbors change the Voronoi volumes of their neighbors, 
and so on. Therefore, adding a new cell does not sample the same distribution as before - the distribution
itself changes, rendering the limit inapplicable.

In the case of the \textit{Volvox}, the somatic cells are originally connected together only by cytoplasmic
bridges, forming a small sphere.  As the ECM is generated the sphere ``inflates''. This process, in which many 
random fluctuations in the amount of ECM excreted by each cell over time can integrate together, seems 
appropriate for CLT-like arguments. However, it is worth noting that the cells are generally locally oriented with a 
hexagonal symmetry. In order to maintain a non-wrinkled surface, more ECM must be secreted in some local regions, 
such as the corners of the hexagons, than in other places, such as at the hexagon edges. Since there is no 
local wrinkling observed, the secretion of ECM from the somatic cells cannot be a completely random process 
orientationally. In other words, the ECM excretion process is controlled, which implies that the CLT does not 
properly capture the sampling space. Instead, the cells inevitably occupy positions on the surface of the sphere 
that vary from organism to organism; the maximum entropy distribution of their Voronoi areas is then the k-gamma distribution.

\bibliography{Day_etal_bib}

\begin{thebibliography}{79}
\providecommand{\natexlab}[1]{#1}
\providecommand{\urlprefix}{}
\providecommand{\doiprefix}{doi: }

\bibitem[{Alsous et~al.(2018)Alsous, Jasmin Imran and Villoutreix, Paul and
  Stoop, Norbert and Shvartsman, Stanislav Y and Dunkel,
  J{\"{o}}rn}]{Alsous2018}
\textbf{\color{eLifeMediumGrey} Alsous JI}, Villoutreix P, Stoop N, Shvartsman
  SY, Dunkel J.
\newblock {Entropic effects in cell lineage tree packings}.
\newblock Nature Physics.  2018;
  \href{http://dx.doi.org/10.1038/s41567-018-0202-0}{\doiprefix
  \detokenize{http://dx.doi.org/10.1038/s41567-018-0202-0}}.

\bibitem[{Angert(2005)Angert, Esther R}]{Angert2005}
\textbf{\color{eLifeMediumGrey} Angert ER}.
\newblock {Alternatives to binary fission in bacteria}.
\newblock Nature Reviews Microbiology.  2005; 3(March):214--224.
\newblock \href{http://dx.doi.org/10.1038/nrmicro1096}{\doiprefix
  \detokenize{http://dx.doi.org/10.1038/nrmicro1096}}.

\bibitem[{Aste and {Di Matteo}(2008)Aste, T. and {Di Matteo}, T.}]{Aste2008}
\textbf{\color{eLifeMediumGrey} Aste T}, {Di Matteo} T.
\newblock {Emergence of Gamma distributions in granular materials and packing
  models}.
\newblock Physical Review E.  2008; 77(2):1--8.
\newblock \href{http://dx.doi.org/10.1103/PhysRevE.77.021309}{\doiprefix
  \detokenize{http://dx.doi.org/10.1103/PhysRevE.77.021309}}.

\bibitem[{Aste et~al.(2007)Aste, T. and {Di Matteo}, T. and Saadatfar, M. and
  Senden, T. J. and Schr{\"{o}}ter, Matthias and Swinney, Harry L.}]{Aste2007}
\textbf{\color{eLifeMediumGrey} Aste T}, {Di Matteo} T, Saadatfar M, Senden TJ,
  Schr{\"{o}}ter M, Swinney HL.
\newblock {An invariant distribution in static granular media}.
\newblock Europhysics Letters.  2007; 79(2).
\newblock \href{http://dx.doi.org/10.1209/0295-5075/79/24003}{\doiprefix
  \detokenize{http://dx.doi.org/10.1209/0295-5075/79/24003}}.

\bibitem[{Atia et~al.(2018)Atia, Lior and Bi, Dapeng and Sharma, Yasha and
  Mitchel, Jennifer A. and Gweon, Bomi and Koehler, Stephan A. and Decamp,
  Stephen J. and Lan, Bo and Kim, Jae Hun and Hirsch, Rebecca and Pegoraro,
  Adrian F. and Lee, Kyu Ha and Starr, Jacqueline R. and Weitz, David A. and
  Martin, Adam C. and Park, Jin Ah and Butler, James P. and Fredberg, Jeffrey
  J.}]{Atia2018}
\textbf{\color{eLifeMediumGrey} Atia L}, Bi D, Sharma Y, Mitchel JA, Gweon B,
  Koehler SA, Decamp SJ, Lan B, Kim JH, Hirsch R, Pegoraro AF, Lee KH, Starr
  JR, Weitz DA, Martin AC, Park JA, Butler JP, Fredberg JJ.
\newblock {Geometric constraints during epithelial jamming}.
\newblock Nature Physics.  2018; 14(6):613--620.
\newblock \href{http://dx.doi.org/10.1038/s41567-018-0089-9}{\doiprefix
  \detokenize{http://dx.doi.org/10.1038/s41567-018-0089-9}}.

\bibitem[{Bell and Mooers(1997)Bell, G. and Mooers, A.O.}]{BellMoers}
\textbf{\color{eLifeMediumGrey} Bell G}, Mooers AO.
\newblock {Size and complexity among multicellular organisms}.
\newblock Biological Journal of the Linnean Society.  1997; 60(3):345--363.
\newblock
  \href{http://dx.doi.org/10.1111/j.1095-8312.1997.tb01500.x}{\doiprefix
  \detokenize{http://dx.doi.org/10.1111/j.1095-8312.1997.tb01500.x}}.

\bibitem[{Bi et~al.(2015{\natexlab{a}})Bi, Dapeng and Henkes, Silke and
  Daniels, Karen E. and Chakraborty, Bulbul}]{Bi2015}
\textbf{\color{eLifeMediumGrey} Bi D}, Henkes S, Daniels KE, Chakraborty B.
\newblock {The statistical physics of athermal materials}.
\newblock Annual Review of Condensed Matter Physics.  2015; 6:63--83.
\newblock
  \href{http://dx.doi.org/10.1146/annurev-conmatphys-031214-014336}{\doiprefix
  \detokenize{http://dx.doi.org/10.1146/annurev-conmatphys-031214-014336}}.

\bibitem[{Bi et~al.(2015{\natexlab{b}})Bi, Dapeng and Lopez, J. H. and Schwarz,
  J. M. and Manning, M. Lisa}]{Bi2015a}
\textbf{\color{eLifeMediumGrey} Bi D}, Lopez JH, Schwarz JM, Manning ML.
\newblock {A density-independent rigidity transition in biological tissues}.
\newblock Nature Physics.  2015; 11(12):1074--1079.
\newblock \href{http://dx.doi.org/10.1038/nphys3471}{\doiprefix
  \detokenize{http://dx.doi.org/10.1038/nphys3471}}.

\bibitem[{Bonner(1998)Bonner, John Tyler}]{Bonner1998}
\textbf{\color{eLifeMediumGrey} Bonner JT}.
\newblock {The origins of multicellularity}.
\newblock Integrative Biology: Issues, News, and Reviews.  1998; 1(1):27--36.
\newblock
  \href{http://dx.doi.org/10.1002/(sici)1520-6602(1998)1:1<27::aid-inbi4>3.3.co;2-y}{\doiprefix
  \detokenize{http://dx.doi.org/10.1002/(sici)1520-6602(1998)1:1<27::aid-inbi4>3.3.co;2-y}}.

\bibitem[{Boraas et~al.(1998)Boraas, Martin E. and Seale, Dianne B. and
  Boxhorn, Joseph E.}]{Boraas1998}
\textbf{\color{eLifeMediumGrey} Boraas ME}, Seale DB, Boxhorn JE.
\newblock {Phagotrophy by flagellate selects for colonial prey: A possible
  origin of multicellularity}.
\newblock Evolutionary Ecology.  1998; 12(2):153--164.
\newblock \href{http://dx.doi.org/10.1023/A:1006527528063}{\doiprefix
  \detokenize{http://dx.doi.org/10.1023/A:1006527528063}}.

\bibitem[{Brochard and Lennon(1975)Brochard, F. and Lennon, J.F.}]{flicker}
\textbf{\color{eLifeMediumGrey} Brochard F}, Lennon JF.
\newblock Frequency spectrum of the flicker phenomenon in erythrocytes.
\newblock Journale de Physique.  1975; 36:1035--1047.
\newblock \href{http://dx.doi.org/10.1051/jphys:0197500360110103500}{\doiprefix
  \detokenize{http://dx.doi.org/10.1051/jphys:0197500360110103500}}.

\bibitem[{Brumley et~al.(2014)Brumley, Douglas R and Wan, Kirsty Y and Polin,
  Marco and Goldstein, Raymond E}]{Brumley2014}
\textbf{\color{eLifeMediumGrey} Brumley DR}, Wan KY, Polin M, Goldstein RE.
\newblock {Flagellar synchronization through direct hydrodynamic interactions}.
\newblock eLife.  2014; 3:1--15.
\newblock \href{http://dx.doi.org/10.7554/elife.02750}{\doiprefix
  \detokenize{http://dx.doi.org/10.7554/elife.02750}}.

\bibitem[{Brumley et~al.(2015)Brumley, D.R. and Polin, M. and Pedley, T.J. and
  Goldstein, R.E.}]{Brumley2015}
\textbf{\color{eLifeMediumGrey} Brumley DR}, Polin M, Pedley TJ, Goldstein RE.
\newblock Metachronal waves in the flagellar beating of {\it Volvox} and their
  hydrodynamic origin.
\newblock Journal of the Royal Society Interface.  2015; 12:20141358.
\newblock \href{http://dx.doi.org/10.1098/rsif.2014.1358}{\doiprefix
  \detokenize{http://dx.doi.org/10.1098/rsif.2014.1358}}.

\bibitem[{Brunet et~al.(2019)Brunet, Thibaut and Larson, Ben T. and Linden,
  Tess A. and Vermeij, Mark J. A. and McDonald, Kent and King,
  Nicole}]{Brunet2019}
\textbf{\color{eLifeMediumGrey} Brunet T}, Larson BT, Linden TA, Vermeij MJA,
  McDonald K, King N.
\newblock {Light-regulated collective contractility in a multicellular
  choanoflagellate}.
\newblock Science.  2019; 366:326--334.
\newblock \href{http://dx.doi.org/10.1126/science.aay2346}{\doiprefix
  \detokenize{http://dx.doi.org/10.1126/science.aay2346}}.

\bibitem[{Butterfield(2000)Butterfield, Nicholas
  J}]{butterfield2000bangiomorpha}
\textbf{\color{eLifeMediumGrey} Butterfield NJ}.
\newblock {\it Bangiomorpha pubescens} n. gen., n. sp.: implications for the
  evolution of sex, multicellularity, and the Mesoproterozoic/Neoproterozoic
  radiation of eukaryotes.
\newblock Paleobiology.  2000; 26(3):386--404.
\newblock
  \href{http://dx.doi.org/10.1666/0094-8373(2000)026<0386:BPNGNS>2.0.CO;2}{\doiprefix
  \detokenize{http://dx.doi.org/10.1666/0094-8373(2000)026<0386:BPNGNS>2.0.CO;2}}.

\bibitem[{Chant and Pringle(1995)Chant, John and Pringle, John R.}]{Chant1995}
\textbf{\color{eLifeMediumGrey} Chant J}, Pringle JR.
\newblock {Patterns of bud-site selection in the yeast Saccharomyces
  cerevisiae}.
\newblock Journal of Cell Biology.  1995; 129(3):751--765.
\newblock \href{http://dx.doi.org/10.1083/jcb.129.3.751}{\doiprefix
  \detokenize{http://dx.doi.org/10.1083/jcb.129.3.751}}.

\bibitem[{Claessen et~al.(2014)Claessen, Dennis and Rozen, Daniel E. and
  Kuipers, Oscar P. and S{\o}gaard-Andersen, Lotte and {Van Wezel}, Gilles
  P.}]{Claessen2014}
\textbf{\color{eLifeMediumGrey} Claessen D}, Rozen DE, Kuipers OP,
  S{\o}gaard-Andersen L, {Van Wezel} GP.
\newblock {Bacterial solutions to multicellularity: A tale of biofilms,
  filaments and fruiting bodies}.
\newblock Nature Reviews Microbiology.  2014; 12(2):115--124.
\newblock \href{http://dx.doi.org/10.1038/nrmicro3178}{\doiprefix
  \detokenize{http://dx.doi.org/10.1038/nrmicro3178}}.

\bibitem[{Colizzi et~al.(2020)Colizzi, Enrico Sandro and Vroomans, Renske M.A.
  and Merks, Roeland M.H.}]{Colizzi2020}
\textbf{\color{eLifeMediumGrey} Colizzi ES}, Vroomans RMA, Merks RMH.
\newblock {Evolution of multicellularity by collective integration of spatial
  information}.
\newblock eLife.  2020; 9:1--57.
\newblock \href{10.7554/eLife.56349}{\doiprefix
  \detokenize{10.7554/eLife.56349}}.

\bibitem[{Damavandi and Lubensky(2019)Damavandi, Ojan Khatib and Lubensky,
  David K.}]{Damavandi2019}
\textbf{\color{eLifeMediumGrey} Damavandi OK}, Lubensky DK.
\newblock {Statistics of noisy growth with mechanical feedback in elastic
  tissues}.
\newblock Proceedings of the National Academy of Sciences.  2019;
  116(12):5350--5355.
\newblock \href{http://dx.doi.org/10.1073/pnas.1816100116}{\doiprefix
  \detokenize{http://dx.doi.org/10.1073/pnas.1816100116}}.

\bibitem[{Davidson(2001)Eric H. Davidson}]{Davidson2001}
\textbf{\color{eLifeMediumGrey} Davidson EH}.
\newblock Chapter 1 - Regulatory Hardwiring: A Brief Overview of the Genomic
  Control Apparatus and its Causal Role in Development and Evolution.
\newblock In: Davidson EH, editor. \emph{Genomic Regulatory Systems} San Diego:
  Academic Press; 2001.p. 1--23.
\newblock \href{https://doi.org/10.1016/B978-012205351-1.50001-X}{\doiprefix
  \detokenize{https://doi.org/10.1016/B978-012205351-1.50001-X}}.

\bibitem[{Delarue et~al.(2016)Delarue, Morgan and Hartung, J{\"{o}}rn and
  Schreck, Carl and Gniewek, Pawel and Hu, Lucy and Herminghaus, Stephan and
  Hallatschek, Oskar}]{Delarue2016}
\textbf{\color{eLifeMediumGrey} Delarue M}, Hartung J, Schreck C, Gniewek P, Hu
  L, Herminghaus S, Hallatschek O.
\newblock {Self-driven jamming in growing microbial populations}.
\newblock Nature Physics.  2016; 12.
\newblock \href{http://dx.doi.org/10.1038/NPHYS3741}{\doiprefix
  \detokenize{http://dx.doi.org/10.1038/NPHYS3741}}.

\bibitem[{Deneke and {Di Talia}(2018)Deneke, Victoria E. and {Di Talia},
  Stefano}]{Deneke2018}
\textbf{\color{eLifeMediumGrey} Deneke VE}, {Di Talia} S.
\newblock {Chemical waves in cell and developmental biology}.
\newblock Journal of Cell Biology.  2018; 217(4):1193--1204.
\newblock \href{http://dx.doi.org/10.1083/jcb.201701158}{\doiprefix
  \detokenize{http://dx.doi.org/10.1083/jcb.201701158}}.

\bibitem[{Drescher et~al.(2010{\natexlab{a}})Drescher, K. and Goldstein, R.E.
  and Michel, N. and Polin, M. and Tuval, I.}]{Drescher}
\textbf{\color{eLifeMediumGrey} Drescher K}, Goldstein RE, Michel N, Polin M,
  Tuval I.
\newblock Direct measurement of the flow field around swimming microorganisms.
\newblock Physical Review Letters.  2010; 105:168101.
\newblock \href{http://dx.doi.org/10.1103/PhysRevLett.105.168101}{\doiprefix
  \detokenize{http://dx.doi.org/10.1103/PhysRevLett.105.168101}}.

\bibitem[{Drescher et~al.(2010{\natexlab{b}})Drescher, K. and Goldstein, R.E.
  and Tuval, I.}]{Fidelity}
\textbf{\color{eLifeMediumGrey} Drescher K}, Goldstein RE, Tuval I.
\newblock Fidelity of adaptive phototaxis.
\newblock Proceedings of the National Academy of Science.  2010;
  107:11171--11176.
\newblock \href{http://dx.doi.org/10.1073/pnas.1000901107}{\doiprefix
  \detokenize{http://dx.doi.org/10.1073/pnas.1000901107}}.

\bibitem[{Drescher et~al.(2016)Drescher, Knut and Dunkel, J{\"{o}}rn and
  Nadell, Carey D. and {Van Teeffelen}, Sven and Grnja, Ivan and Wingreen, Ned
  S. and Stone, Howard A. and Bassler, Bonnie L.}]{Drescher2016}
\textbf{\color{eLifeMediumGrey} Drescher K}, Dunkel J, Nadell CD, {Van
  Teeffelen} S, Grnja I, Wingreen NS, Stone HA, Bassler BL.
\newblock {Architectural transitions in Vibrio cholerae biofilms at single-cell
  resolution}.
\newblock Proceedings of the National Academy of Sciences.  2016;
  113(14):E2066--E2072.
\newblock \href{http://dx.doi.org/10.1073/pnas.1601702113}{\doiprefix
  \detokenize{http://dx.doi.org/10.1073/pnas.1601702113}}.

\bibitem[{Fisher et~al.(2016)Fisher, R M and Bell, T and West, S
  A}]{Fisher2016}
\textbf{\color{eLifeMediumGrey} Fisher RM}, Bell T, West SA.
\newblock {Multicellular group formation in response to predators in the alga
  {\it Chlorella vulgaris}}.
\newblock Journal of Evolutionary Biology.  2016; 29:551--559.
\newblock \href{http://dx.doi.org/10.1111/jeb.12804}{\doiprefix
  \detokenize{http://dx.doi.org/10.1111/jeb.12804}}.

\bibitem[{Goldstein(2015)Goldstein, R.E.}]{GoldsteinARFM}
\textbf{\color{eLifeMediumGrey} Goldstein RE}.
\newblock {Green algae as model organisms for biological fluid dynamics}.
\newblock Annual Review of Fluid Mechanics.  2015; 47:343--375.
\newblock
  \href{http://dx.doi.org/10.1146/annurev-fluid-010313-141426}{\doiprefix
  \detokenize{http://dx.doi.org/10.1146/annurev-fluid-010313-141426}}.

\bibitem[{Gregor et~al.(2007)Gregor, Thomas and Tank, David W. and Wieschaus,
  Eric F. and Bialek, William}]{Gregor2007}
\textbf{\color{eLifeMediumGrey} Gregor T}, Tank DW, Wieschaus EF, Bialek W.
\newblock {Probing the Limits to Positional Information}.
\newblock Cell.  2007; 130(1):153--164.
\newblock \href{http://dx.doi.org/10.1016/j.cell.2007.05.025}{\doiprefix
  \detokenize{http://dx.doi.org/10.1016/j.cell.2007.05.025}}.

\bibitem[{Grosberg and Strathmann(2007)Grosberg, Richard K. and Strathmann,
  Richard R.}]{Grosberg2007}
\textbf{\color{eLifeMediumGrey} Grosberg RK}, Strathmann RR.
\newblock {The Evolution of Multicellularity: A Minor Major Transition?}
\newblock Annual Review of Ecology, Evolution, and Systematics.  2007;
  38(1):621--654.
\newblock
  \href{http://dx.doi.org/10.1146/annurev.ecolsys.36.102403.114735}{\doiprefix
  \detokenize{http://dx.doi.org/10.1146/annurev.ecolsys.36.102403.114735}}.

\bibitem[{Haas et~al.(2018)Haas, Pierre A. and H{\"{o}}hn, Stephanie S.M.H. and
  Honerkamp-Smith, Aurelia R. and Kirkegaard, Julius B. and Goldstein, Raymond
  E.}]{Haas2018}
\textbf{\color{eLifeMediumGrey} Haas PA}, H{\"{o}}hn SSMH, Honerkamp-Smith AR,
  Kirkegaard JB, Goldstein RE.
\newblock {The noisy basis of morphogenesis: Mechanisms and mechanics of cell
  sheet folding inferred from developmental variability}.
\newblock PLoS Biology.  2018; 16(7):1--37.
\newblock \href{http://dx.doi.org/10.1371/journal.pbio.2005536}{\doiprefix
  \detokenize{http://dx.doi.org/10.1371/journal.pbio.2005536}}.

\bibitem[{Hartmann et~al.(2019)Hartmann, Raimo and Singh, Praveen K. and
  Pearce, Philip and Mok, Rachel and Song, Boya and D{\'{i}}az-Pascual,
  Francisco and Dunkel, J{\"{o}}rn and Drescher, Knut}]{Hartmann2019}
\textbf{\color{eLifeMediumGrey} Hartmann R}, Singh PK, Pearce P, Mok R, Song B,
  D{\'{i}}az-Pascual F, Dunkel J, Drescher K.
\newblock {Emergence of three-dimensional order and structure in growing
  biofilms}.
\newblock Nature Physics.  2019; 15(3):251--256.
\newblock \href{http://dx.doi.org/10.1038/s41567-018-0356-9}{\doiprefix
  \detokenize{http://dx.doi.org/10.1038/s41567-018-0356-9}}.

\bibitem[{Herron et~al.(2019)Herron, Matthew D. and Borin, Joshua M. and
  Boswell, Jacob C. and Walker, Jillian and Chen, I. Chen Kimberly and Knox,
  Charles A. and Boyd, Margrethe and Rosenzweig, Frank and Ratcliff, William
  C.}]{Herron2019}
\textbf{\color{eLifeMediumGrey} Herron MD}, Borin JM, Boswell JC, Walker J,
  Chen ICK, Knox CA, Boyd M, Rosenzweig F, Ratcliff WC.
\newblock {De novo origins of multicellularity in response to predation}.
\newblock Scientific Reports.  2019; 9(1):1--9.
\newblock \href{http://dx.doi.org/10.1038/s41598-019-39558-8}{\doiprefix
  \detokenize{http://dx.doi.org/10.1038/s41598-019-39558-8}}.

\bibitem[{Herron et~al.(2009)Herron, Matthew D and Hackett, Jeremiah D and
  Aylward, Frank O and Michod, Richard E}]{herron2009triassic}
\textbf{\color{eLifeMediumGrey} Herron MD}, Hackett JD, Aylward FO, Michod RE.
\newblock Triassic origin and early radiation of multicellular volvocine algae.
\newblock Proceedings of the National Academy of Sciences.  2009;
  106(9):3254--3258.
\newblock \href{http://dx.doi.org/10.1073/pnas.0811205106}{\doiprefix
  \detokenize{http://dx.doi.org/10.1073/pnas.0811205106}}.

\bibitem[{Hohenberg(1967)Hohenberg, P. C.}]{Hohenberg1967b}
\textbf{\color{eLifeMediumGrey} Hohenberg PC}.
\newblock {Existence of long-range order in one and two dimensions}.
\newblock Physical Review.  1967; 158(2):383--386.
\newblock \href{http://dx.doi.org/10.1103/PhysRev.158.383}{\doiprefix
  \detokenize{http://dx.doi.org/10.1103/PhysRev.158.383}}.

\bibitem[{Hong et~al.(2016)Hong, Lilan and Dumond, Mathilde and Tsugawa, Satoru
  and Sapala, Aleksandra and Routier-Kierzkowska, Anne Lise and Zhou, Yong and
  Chen, Catherine and Kiss, Annamaria and Zhu, Mingyuan and Hamant, Olivier and
  Smith, Richard S. and Komatsuzaki, Tamiki and Li, Chun Biu and Boudaoud,
  Arezki and Roeder, Adrienne H.K.}]{Hong2016}
\textbf{\color{eLifeMediumGrey} Hong L}, Dumond M, Tsugawa S, Sapala A,
  Routier-Kierzkowska AL, Zhou Y, Chen C, Kiss A, Zhu M, Hamant O, Smith RS,
  Komatsuzaki T, Li CB, Boudaoud A, Roeder AHK.
\newblock {Variable Cell Growth Yields Reproducible OrganDevelopment through
  Spatiotemporal Averaging}.
\newblock Developmental Cell.  2016; 38(1):15--32.
\newblock \href{http://dx.doi.org/10.1016/j.devcel.2016.06.016}{\doiprefix
  \detokenize{http://dx.doi.org/10.1016/j.devcel.2016.06.016}}.

\bibitem[{Ishikawa et~al.(2020)Ishikawa, T. and Pedley, T.J. and Drescher, K.
  and Goldstein, R.E.}]{Ishikawa}
\textbf{\color{eLifeMediumGrey} Ishikawa T}, Pedley TJ, Drescher K, Goldstein
  RE.
\newblock Stability of dancing {\it Volvox}.
\newblock Journal of Fluid Mechanics.  2020; 903:A11.
\newblock \href{http://dx.doi.org/10.1017/jfm.2020.613}{\doiprefix
  \detokenize{http://dx.doi.org/10.1017/jfm.2020.613}}.

\bibitem[{Jacobeen et~al.(2018{\natexlab{a}})Jacobeen, Shane and Graba, Elyes
  C. and Brandys, Colin G. and Day, Thomas C. and Ratcliff, William C. and
  Yunker, Peter J.}]{Jacobeen2018}
\textbf{\color{eLifeMediumGrey} Jacobeen S}, Graba EC, Brandys CG, Day TC,
  Ratcliff WC, Yunker PJ.
\newblock {Geometry, packing, and evolutionary paths to increased multicellular
  size}.
\newblock Physical Review E.  2018; 97(5):1--6.
\newblock \href{http://dx.doi.org/10.1103/PhysRevE.97.050401}{\doiprefix
  \detokenize{http://dx.doi.org/10.1103/PhysRevE.97.050401}}.

\bibitem[{Jacobeen et~al.(2018{\natexlab{b}})Jacobeen, Shane and Pentz,
  Jennifer T and Graba, Elyes C and Brandys, Colin G and Ratcliff, William C
  and Yunker, Peter J}]{Jacobeen2018a}
\textbf{\color{eLifeMediumGrey} Jacobeen S}, Pentz JT, Graba EC, Brandys CG,
  Ratcliff WC, Yunker PJ.
\newblock {Cellular packing , mechanical stress and the evolution of
  multicellularity}.
\newblock Nature Physics.  2018; 14(March):286--291.
\newblock \href{http://dx.doi.org/10.1038/s41567-017-0002-y}{\doiprefix
  \detokenize{http://dx.doi.org/10.1038/s41567-017-0002-y}}.

\bibitem[{Kalziqi et~al.(2018)Kalziqi, Arben and Yanni, David and Thomas, Jacob
  and Ng, Siu Lung and Vivek, Skanda and Hammer, Brian K and Yunker, Peter
  J}]{Kalziqi2018}
\textbf{\color{eLifeMediumGrey} Kalziqi A}, Yanni D, Thomas J, Ng SL, Vivek S,
  Hammer BK, Yunker PJ.
\newblock {Immotile Active Matter : Activity from Death and Reproduction}.
\newblock Physical Review Letters.  2018; 120(1):18101.
\newblock \href{10.1103/PhysRevLett.120.018101}{\doiprefix
  \detokenize{10.1103/PhysRevLett.120.018101}}.

\bibitem[{Katgert and {Van Hecke}(2010)Katgert, G. and {Van Hecke},
  M.}]{Katgert2010}
\textbf{\color{eLifeMediumGrey} Katgert G}, {Van Hecke} M.
\newblock {Jamming and geometry of two-dimensional foams}.
\newblock Europhysics Letters.  2010; 92(3).
\newblock \href{http://dx.doi.org/10.1209/0295-5075/92/34002}{\doiprefix
  \detokenize{http://dx.doi.org/10.1209/0295-5075/92/34002}}.

\bibitem[{Keim et~al.(2004)Keim, Carolina N and Martins, Juliana L and Abreu,
  Fernanda and Soares, Alexandre and Lins, Henrique and Barros, De and
  Borojevic, Radovan and Lins, Ulysses and Farina, Marcos}]{Keim2004}
\textbf{\color{eLifeMediumGrey} Keim CN}, Martins JL, Abreu F, Soares A, Lins
  H, Barros D, Borojevic R, Lins U, Farina M.
\newblock {Multicellular life cycle of magnetotactic prokaryotes}.
\newblock FEMS Microbiology Letters.  2004; 240:203--208.
\newblock \href{http://dx.doi.org/10.1016/j.femsle.2004.09.035}{\doiprefix
  \detokenize{http://dx.doi.org/10.1016/j.femsle.2004.09.035}}.

\bibitem[{Kirk(2005)Kirk, David L.}]{Kirk2005}
\textbf{\color{eLifeMediumGrey} Kirk DL}.
\newblock {A twelve-step program for evolving multicellularity and a division
  of labor}.
\newblock BioEssays.  2005; 27(3):299--310.
\newblock \href{http://dx.doi.org/10.1002/bies.20197}{\doiprefix
  \detokenize{http://dx.doi.org/10.1002/bies.20197}}.

\bibitem[{Knoll(2011)Knoll, Andrew H.}]{Knoll2011}
\textbf{\color{eLifeMediumGrey} Knoll AH}.
\newblock {The Multiple Origins of Complex Multicellularity}.
\newblock Annual Review of Earth and Planetary Sciences.  2011; 39(1):217--239.
\newblock
  \href{http://dx.doi.org/10.1146/annurev.earth.031208.100209}{\doiprefix
  \detokenize{http://dx.doi.org/10.1146/annurev.earth.031208.100209}}.

\bibitem[{Koyama et~al.(1977)Koyama, T. and Yamada, M. and Matsuhashi,
  M.}]{Koyama1977}
\textbf{\color{eLifeMediumGrey} Koyama T}, Yamada M, Matsuhashi M.
\newblock {Formation of regular packets of {\it Staphylococcus aureus} cells}.
\newblock Journal of Bacteriology.  1977; 129(3):1518--1523.
\newblock \href{http://dx.doi.org/10.1128/jb.129.3.1518-1523.1977}{\doiprefix
  \detokenize{http://dx.doi.org/10.1128/jb.129.3.1518-1523.1977}}.

\bibitem[{Lander(2011)Lander, Arthur D}]{Lander2011}
\textbf{\color{eLifeMediumGrey} Lander AD}.
\newblock {Pattern, growth and control}.
\newblock Cell.  2011; 144(6):955--969.
\newblock \href{http://dx.doi.org/10.1016/j.cell.2011.03.009}{\doiprefix
  \detokenize{http://dx.doi.org/10.1016/j.cell.2011.03.009}}.

\bibitem[{Larson et~al.(2019)Larson, Ben T and Ruiz-Herrero, Teresa and Lee,
  Stacey and Kumar, Sanjay and Mahadevan, L and King, Nicole}]{Larson2019}
\textbf{\color{eLifeMediumGrey} Larson BT}, Ruiz-Herrero T, Lee S, Kumar S,
  Mahadevan L, King N.
\newblock {Biophysical principles of choanoflagellate self-organization}.
\newblock Proceedings of the National Academy of Sciences.  2019;
  117(3):1303--1311.
\newblock \href{http://dx.doi.org/10.1073/pnas.1909447117}{\doiprefix
  \detokenize{http://dx.doi.org/10.1073/pnas.1909447117}}.

\bibitem[{Lauga(2020)Lauga, E.}]{Lauga}
\textbf{\color{eLifeMediumGrey} Lauga E}.
\newblock The Fluid Dynamics of Cell Motility.
\newblock Cambridge University Press; 2020.

\bibitem[{Levin(2004)Levin, Michael}]{Levin2004}
\textbf{\color{eLifeMediumGrey} Levin M}.
\newblock {The embryonic origins of left-right asymmetry}.
\newblock Critial Reviews of Oral Biology and Medicine.  2004; 15(4):197--206.
\newblock \href{http://dx.doi.org/10.1177/154411130401500403}{\doiprefix
  \detokenize{http://dx.doi.org/10.1177/154411130401500403}}.

\bibitem[{Lighthill(1952)Lighthill, M.J.}]{Lighthill}
\textbf{\color{eLifeMediumGrey} Lighthill MJ}.
\newblock On the squirming motion of nearly spherical deformable bodies through
  liquids at very small reynolds numbers.
\newblock Communications on Pure and Applied Mathematics.  1952; 5:109--118.
\newblock \href{http://dx.doi.org/10.1002/cpa.3160050201}{\doiprefix
  \detokenize{http://dx.doi.org/10.1002/cpa.3160050201}}.

\bibitem[{Lurling and {Van Donk}(1997)Lurling, Miquel and {Van Donk},
  Ellen}]{Lurling1997}
\textbf{\color{eLifeMediumGrey} Lurling M}, {Van Donk} E.
\newblock {Morphological changes in {\it Scenedesmus} induced by infochemicals
  released in situ from zooplankton grazers}.
\newblock Limnology {\&} Oceanography.  1997; 42(4):783--788.
\newblock \href{http://dx.doi.org/10.4319/lo.1997.42.4.0783}{\doiprefix
  \detokenize{http://dx.doi.org/10.4319/lo.1997.42.4.0783}}.

\bibitem[{Magar et~al.(2003)Magar, V. and Goto, T. and Pedley, T.J.}]{Magar}
\textbf{\color{eLifeMediumGrey} Magar V}, Goto T, Pedley TJ.
\newblock Nutrient Uptake by a Self-Propelled Steady Squirmer.
\newblock Quartely Journal of Mechanics and Applied Mathematics.  2003;
  56:65--91.
\newblock \href{http://dx.doi.org/10.1093/qjmam/56.1.65}{\doiprefix
  \detokenize{http://dx.doi.org/10.1093/qjmam/56.1.65}}.

\bibitem[{Manoharan(2015)Manoharan, Vinothan N.}]{Manoharan2015}
\textbf{\color{eLifeMediumGrey} Manoharan VN}.
\newblock {Colloidal matter: Packing, geometry, and entropy}.
\newblock Science.  2015; 349(6251).
\newblock \href{http://dx.doi.org/10.1126/science.1253751}{\doiprefix
  \detokenize{http://dx.doi.org/10.1126/science.1253751}}.

\bibitem[{Mermin and Wagner(1966)Mermin, N. D. and Wagner, H.}]{Mermin1966}
\textbf{\color{eLifeMediumGrey} Mermin ND}, Wagner H.
\newblock {Absence of ferromagnetism or antiferromagnetism in one- or
  two-dimensional isotropic Heisenberg models}.
\newblock Physical Review Letters.  1966; 17(22):1133--1136.
\newblock \href{http://dx.doi.org/10.1103/PhysRevLett.17.1133}{\doiprefix
  \detokenize{http://dx.doi.org/10.1103/PhysRevLett.17.1133}}.

\bibitem[{Michel and Yunker(2019)Michel, Jonathan A and Yunker, Peter
  J}]{Michel2019}
\textbf{\color{eLifeMediumGrey} Michel JA}, Yunker PJ.
\newblock {Structural hierarchy confers error tolerance in biological
  materials}.
\newblock Proceedings of the National Academy of Science.  2019;
  116(8):2875--2880.
\newblock \href{http://dx.doi.org/10.1073/pnas.1813801116}{\doiprefix
  \detokenize{http://dx.doi.org/10.1073/pnas.1813801116}}.

\bibitem[{Pentz et~al.(2020)Pentz, Jennifer T. and
  M{\'{a}}rquez-Zacar{\'{i}}as, Pedro and Bozdag, G. Ozan and Burnetti, Anthony
  and Yunker, Peter J. and Libby, Eric and Ratcliff, William C.}]{Pentz2020}
\textbf{\color{eLifeMediumGrey} Pentz JT}, M{\'{a}}rquez-Zacar{\'{i}}as P,
  Bozdag GO, Burnetti A, Yunker PJ, Libby E, Ratcliff WC.
\newblock {Ecological Advantages and Evolutionary Limitations of Aggregative
  Multicellular Development}.
\newblock Current Biology.  2020; 30(21):4155--4164.e6.
\newblock \href{http://dx.doi.org/10.1016/j.cub.2020.08.006}{\doiprefix
  \detokenize{http://dx.doi.org/10.1016/j.cub.2020.08.006}}.

\bibitem[{Prakash et~al.(2019)Prakash, Vivek N. and Bull, Matthew S. and
  Prakash, Manu}]{Prakash2019}
\textbf{\color{eLifeMediumGrey} Prakash VN}, Bull MS, Prakash M.
\newblock {Motility induced fracture reveals a ductile-to-brittle crossover in
  a simple animal's epithelial}.
\newblock Nature Physics.  2019; 17:504--511.
\newblock \href{http://dx.doi.org/10.1038/s41567-020-01134-7}{\doiprefix
  \detokenize{http://dx.doi.org/10.1038/s41567-020-01134-7}}.

\bibitem[{Ratcliff et~al.(2012)Ratcliff, William C. and Denison, R. Ford and
  Borrello, Mark and Travisano, Michael}]{Ratcliff2012}
\textbf{\color{eLifeMediumGrey} Ratcliff WC}, Denison RF, Borrello M, Travisano
  M.
\newblock {Experimental evolution of multicellularity}.
\newblock Proceedings of the National Academy of Sciences.  2012;
  109(5):1595--1600.
\newblock \href{http://dx.doi.org/10.1073/pnas.1115323109}{\doiprefix
  \detokenize{http://dx.doi.org/10.1073/pnas.1115323109}}.

\bibitem[{Ratcliff et~al.(2015)Ratcliff, William C. and Fankhauser, Johnathon
  D. and Rogers, David W. and Greig, Duncan and Travisano,
  Michael}]{Ratcliff2015}
\textbf{\color{eLifeMediumGrey} Ratcliff WC}, Fankhauser JD, Rogers DW, Greig
  D, Travisano M.
\newblock {Origins of multicellular evolvability in snowflake yeast}.
\newblock Nature Communications.  2015; 6(May 2014):1--9.
\newblock \href{http://dx.doi.org/10.1038/ncomms7102}{\doiprefix
  \detokenize{http://dx.doi.org/10.1038/ncomms7102}}.

\bibitem[{Ratcliff et~al.(2013)Ratcliff, William C and Herron, Matthew D and
  Howell, Kathryn and Pentz, Jennifer T and Rosenzweig, Frank and Travisano,
  Michael}]{Ratcliff2013}
\textbf{\color{eLifeMediumGrey} Ratcliff WC}, Herron MD, Howell K, Pentz JT,
  Rosenzweig F, Travisano M.
\newblock {Experimental evolution of an alternating uni- and multicellular life
  cycle in {\it Chlamydomonas reinhardtii}}.
\newblock Nature Communications.  2013; 4(May).
\newblock \href{http://dx.doi.org/10.1038/ncomms3742}{\doiprefix
  \detokenize{http://dx.doi.org/10.1038/ncomms3742}}.

\bibitem[{Risler et~al.(2015)Risler, T. and Peilloux, A. and Prost,
  J.}]{Risler}
\textbf{\color{eLifeMediumGrey} Risler T}, Peilloux A, Prost J.
\newblock Homeostatic Fluctuations of a Tissue Surface.
\newblock Physical Review Letters.  2015; 115:258104.
\newblock \href{http://dx.doi.org/10.1103/PhysRevLett.115.258104}{\doiprefix
  \detokenize{http://dx.doi.org/10.1103/PhysRevLett.115.258104}}.

\bibitem[{Rycroft(2009)Rycroft, Chris H}]{Rycroft2009}
\textbf{\color{eLifeMediumGrey} Rycroft CH}.
\newblock {VORO ++ : A three-dimensional Voronoi cell library in C ++}.
\newblock Chaos.  2009; 19(041111).
\newblock \href{http://dx.doi.org/10.1063/1.3215722}{\doiprefix
  \detokenize{http://dx.doi.org/10.1063/1.3215722}}.

\bibitem[{Sampathkumar(2020)Sampathkumar, Arun}]{Sampathkumar2020}
\textbf{\color{eLifeMediumGrey} Sampathkumar A}.
\newblock {Mechanical feedback-loop regulation of morphogenesis in plants}.
\newblock Development.  2020; 147(16).
\newblock \href{http://dx.doi.org/10.1242/dev.177964}{\doiprefix
  \detokenize{http://dx.doi.org/10.1242/dev.177964}}.

\bibitem[{Schmideder et~al.(2021)Schmideder, Stefan and M{\"{u}}ller, Henri and
  Barthel, Lars and Niessen, Ludwig and Meyer, Vera and Briesen,
  Heiko}]{Schmideder2021}
\textbf{\color{eLifeMediumGrey} Schmideder S}, M{\"{u}}ller H, Barthel L,
  Niessen L, Meyer V, Briesen H.
\newblock {Universal law for diffusive mass transport through mycelial
  networks}.
\newblock Biotechnology and Bioengineering.  2021; 118:930--943.
\newblock \href{10.1002/bit.27622}{\doiprefix \detokenize{10.1002/bit.27622}}.

\bibitem[{Schneider et~al.(1984)Schneider, M.B. and Jenkins, J.T. and Webb,
  W.W.}]{SchneiderJenkinsWebb}
\textbf{\color{eLifeMediumGrey} Schneider MB}, Jenkins JT, Webb WW.
\newblock Thermal fluctuations of large quasi-spherical bimolecular
  phospholipid vesicles.
\newblock Biophysical Journal.  1984; 45:891--899.
\newblock \href{http://dx.doi.org/10.1016/S0006-3495(84)84235-6}{\doiprefix
  \detokenize{http://dx.doi.org/10.1016/S0006-3495(84)84235-6}}.

\bibitem[{Shinbrot and Muzzio(2001)Shinbrot, T. and Muzzio, F.
  J.}]{Shinbrot2001}
\textbf{\color{eLifeMediumGrey} Shinbrot T}, Muzzio FJ.
\newblock {Noise to order}.
\newblock Nature.  2001; 410(6825):251--258.
\newblock \href{http://dx.doi.org/10.1038/35065689}{\doiprefix
  \detokenize{http://dx.doi.org/10.1038/35065689}}.

\bibitem[{Short et~al.(2006)Short, M.B. and Solari, C.A. and Ganguly, S. and
  Powers, T.R. and Kessler, J.O. and Goldstein, R.E.}]{Short}
\textbf{\color{eLifeMediumGrey} Short MB}, Solari CA, Ganguly S, Powers TR,
  Kessler JO, Goldstein RE.
\newblock Flows driven by flagella of multicellular organisms enhance
  long-range molecular transport.
\newblock Proceedings of the National Academy of Science.  2006;
  1103:8315--8319.
\newblock \href{http://dx.doi.org/10.1073/pnas.0600566103}{\doiprefix
  \detokenize{http://dx.doi.org/10.1073/pnas.0600566103}}.

\bibitem[{Snoeijer et~al.(2004)Snoeijer, Jacco H. and Vlugt, Thijs J.H. and van
  Hecke, Martin and van Saarloos, Wim}]{Snoeijer2004}
\textbf{\color{eLifeMediumGrey} Snoeijer JH}, Vlugt TJH, van Hecke M, van
  Saarloos W.
\newblock {Force Network Ensemble: A New Approach to Static Granular Matter}.
\newblock Physical Review Letters.  2004; 92(5):4.
\newblock \href{http://dx.doi.org/10.1103/PhysRevLett.92.054302}{\doiprefix
  \detokenize{http://dx.doi.org/10.1103/PhysRevLett.92.054302}}.

\bibitem[{Starr(1969)Starr, R.C.}]{Starr1969}
\textbf{\color{eLifeMediumGrey} Starr RC}.
\newblock Structure, reproduction and differentiation in Volvox carteri f.
  nagariensis lyengar, strains HK9 and 10.
\newblock Archiv f{\"u}r Protistenkunde.  1969; 111:204--222.

\bibitem[{Stratford(1992)Stratford, Malcolm}]{Stratford1992}
\textbf{\color{eLifeMediumGrey} Stratford M}.
\newblock {Yeast Flocculation : A New Perspective}.
\newblock Advances in Microbial Physiology.  1992; 33.
\newblock \href{http://dx.doi.org/10.1016/S0065-2911(08)60215-5}{\doiprefix
  \detokenize{http://dx.doi.org/10.1016/S0065-2911(08)60215-5}}.

\bibitem[{Szavits-Nossan et~al.(2014)Szavits-Nossan, Juraj and Eden, Kym and
  Morris, Ryan J. and Macphee, Cait E. and Evans, Martin R. and Allen, Rosalind
  J.}]{Szavits-Nossan2014}
\textbf{\color{eLifeMediumGrey} Szavits-Nossan J}, Eden K, Morris RJ, Macphee
  CE, Evans MR, Allen RJ.
\newblock {Inherent variability in the kinetics of autocatalytic protein
  self-assembly}.
\newblock Physical Review Letters.  2014; 113(9):1--5.
\newblock \href{http://dx.doi.org/10.1103/PhysRevLett.113.098101}{\doiprefix
  \detokenize{http://dx.doi.org/10.1103/PhysRevLett.113.098101}}.

\bibitem[{Tang et~al.(2020)Tang, Qing and Pang, Ke and Yuan, Xunlai and Xiao,
  Shuhai}]{tang2020one}
\textbf{\color{eLifeMediumGrey} Tang Q}, Pang K, Yuan X, Xiao S.
\newblock A one-billion-year-old multicellular chlorophyte.
\newblock Nature Ecology \& Evolution.  2020; 4(4):543--549.
\newblock \href{http://dx.doi.org/10.1038/s41559-020-1122-9}{\doiprefix
  \detokenize{http://dx.doi.org/10.1038/s41559-020-1122-9}}.

\bibitem[{Tsimring(2014)Tsimring, Lev S.}]{Tsimring2014}
\textbf{\color{eLifeMediumGrey} Tsimring LS}.
\newblock {Noise in biology}.
\newblock Reports on Progress in Physics.  2014; 77(2).
\newblock \href{http://dx.doi.org/10.1088/0034-4885/77/2/026601}{\doiprefix
  \detokenize{http://dx.doi.org/10.1088/0034-4885/77/2/026601}}.

\bibitem[{Varadan and Solomon(2003)Varadan, Priya and Solomon, Michael
  J.}]{Varadan2003}
\textbf{\color{eLifeMediumGrey} Varadan P}, Solomon MJ.
\newblock {Direct visualization of long-range heterogeneous structure in dense
  colloidal gels}.
\newblock Langmuir.  2003; 19(3):509--512.
\newblock \href{http://dx.doi.org/10.1021/la026303j}{\doiprefix
  \detokenize{http://dx.doi.org/10.1021/la026303j}}.

\bibitem[{Vivek et~al.(2017)Vivek, Skanda and Kelleher, Colm P. and Chaikin,
  Paul M. and Weeks, Eric R.}]{Vivek2017a}
\textbf{\color{eLifeMediumGrey} Vivek S}, Kelleher CP, Chaikin PM, Weeks ER.
\newblock {Long-wavelength fluctuations and the glass transition in two
  dimensions and three dimensions}.
\newblock Proceedings of the National Academy of Sciences.  2017;
  114(8):1850--1855.
\newblock \href{http://dx.doi.org/10.1073/pnas.1607226113}{\doiprefix
  \detokenize{http://dx.doi.org/10.1073/pnas.1607226113}}.

\bibitem[{Waddington(1957)Waddington, C.H.}]{Waddington2014}
\textbf{\color{eLifeMediumGrey} Waddington CH}.
\newblock The Strategy of the Genes.
\newblock Routledge; 1957.

\bibitem[{Xiao et~al.(1998)Xiao, Shuhai and Zhang, Yun and Knoll, Andrew
  H}]{Xiao1998}
\textbf{\color{eLifeMediumGrey} Xiao S}, Zhang Y, Knoll AH.
\newblock {Three-dimensional preservation of algae and animal embryos in a
  Neoproterozoic phosphorite}.
\newblock Nature.  1998; 391(February):553--558.
\newblock \href{http://dx.doi.org/10.1038/35318}{\doiprefix
  \detokenize{http://dx.doi.org/10.1038/35318}}.

\bibitem[{Yanni et~al.(2020)Yanni, David and Jacobeen, Shane and
  M{\'a}rquez-Zacar{\'\i}as, Pedro and Weitz, Joshua S and Ratcliff, William C
  and Yunker, Peter J}]{yanni2020topological}
\textbf{\color{eLifeMediumGrey} Yanni D}, Jacobeen S, M{\'a}rquez-Zacar{\'\i}as
  P, Weitz JS, Ratcliff WC, Yunker PJ.
\newblock Topological constraints in early multicellularity favor reproductive
  division of labor.
\newblock eLife.  2020; 9:e54348.
\newblock \href{http://dx.doi.org/10.7554/eLife.54348}{\doiprefix
  \detokenize{http://dx.doi.org/10.7554/eLife.54348}}.

\bibitem[{Zamani-Dahaj et~al.(2021)Zamani-Dahaj, Seyed Alireza and Burnetti,
  Anthony and Day, Thomas C. and Yunker, Peter J. and Ratcliff, William C. and
  Herron, Matthew D.}]{Zamani-Dahaj2021.07.19.452990}
\textbf{\color{eLifeMediumGrey} Zamani-Dahaj SA}, Burnetti A, Day TC, Yunker
  PJ, Ratcliff WC, Herron MD.
\newblock Spontaneous emergence of multicellular heritability.
\newblock bioRxiv.  2021;
  \urlprefix\url{https://www.biorxiv.org/content/early/2021/07/20/2021.07.19.452990.1},
  \href{10.1101/2021.07.19.452990}{\doiprefix
  \detokenize{10.1101/2021.07.19.452990}}.

\bibitem[{Zeravcic and Brenner(2014)Zeravcic, Zorana and Brenner, Michael
  P.}]{Zeravcic2014}
\textbf{\color{eLifeMediumGrey} Zeravcic Z}, Brenner MP.
\newblock {Self-replicating colloidal clusters}.
\newblock Proceedings of the National Academy of Sciences.  2014;
  111(5):1748--1753.
\newblock \href{http://dx.doi.org/10.1073/pnas.1313601111}{\doiprefix
  \detokenize{http://dx.doi.org/10.1073/pnas.1313601111}}.

\end{thebibliography}

\end{document}